\documentclass[10pt]{article}

\usepackage{graphicx}
\usepackage{graphics}
\usepackage{amscd}
\usepackage{amsmath}
\usepackage{amsfonts}
\usepackage{amssymb}
\usepackage{color}

\DeclareMathOperator{\wt}{wt_H}

\newcommand{\bF}{ {\mathbb F}}

\newtheorem{theorem}{Theorem}

\newtheorem{example}[theorem]{Example}

\newtheorem{lemma}[theorem]{Lemma}

\newtheorem{proposition}[theorem]{Proposition}
\newtheorem{remark}[theorem]{Remark}

\begin{document}

\title{Two classes of $p$-ary linear codes and their duals\footnote{
 *Corresponding author.\,\, E-Mail addresses:
 waxiqq@163.com (X. Wang),
 dzheng@hubu.edu.cn(D. Zheng), Zhangyan@hubu.edu.cn(Y. Zhang)}}

\author{ Xiaoqiang Wang$^1$, Dabin Zheng*$^1$ and Yan Zhang$^2$}

\date{\small 1. Hubei Province Key Laboratory of Applied Mathematics, \\
Faculty of Mathematics and Statistics, Hubei University, Wuhan 430062, China \\
2. School of Computer Science and Information Engineering, Hubei University, Wuhan 430062, China
}
\maketitle

\leftskip 1.0in
\rightskip 1.0in
\noindent {\bf Abstract.} Let $\mathbb{F}_{p^m}$ be the finite field of order $p^m$, where $p$ is an odd prime and $m$ is a positive integer. In this paper, we investigate a class of subfield codes of linear codes and obtain the weight distribution of
\begin{equation*}
\begin{split}
\mathcal{C}_k=\left\{\left(\left( {\rm Tr}_1^m\left(ax^{p^k+1}+bx\right)+c\right)_{x \in \mathbb{F}_{p^m}}, {\rm Tr}_1^m(a)\right) : \, a,b \in \mathbb{F}_{p^m}, c \in \mathbb{F}_p\right\},
\end{split}
\end{equation*}
 where $k$ is a nonnegative integer.  Our results generalize the results of the subfield codes of the conic codes in \cite{Hengar}.  Among other results, we study the punctured code of $\mathcal{C}_k$, which is defined as
$$\mathcal{\bar{C}}_k=\left\{\left( {\rm Tr}_1^m\left(a x^{{p^k}+1}+bx\right)+c\right)_{x \in \mathbb{F}_{p^m}} : \, a,b \in \mathbb{F}_{p^m}, \,\,c \in \mathbb{F}_p\right\}.$$
The parameters of these linear codes are new in some cases. Some of the presented codes are optimal or almost optimal. Moreover, let $v_2(\cdot)$ denote the 2-adic order function and $v_2(0)=\infty$, the duals of $\mathcal{C}_k$ and $\mathcal{\bar{C}}_k$ are optimal with respect to the Sphere Packing bound if $p>3$, and the dual of $\mathcal{\bar{C}}_k$ is an optimal ternary linear code for the case $v_2(m)\leq v_2(k)$ if $p=3$ and $m>1$.


\vskip 35pt

\leftskip 0.0in
\rightskip 0.0in

\section{Introduction}
Let $p$ be an odd prime and $\mathbb{F}_{p^m}$ be a finite field of size $p^m$.
 An $[n, k,d]$ code $\mathcal{C}$ over the finite field $\mathbb{F}_{p^m}$ is a $k$-dimensional linear subspace of $\mathbb{F}_{p^m}^n$ with the minimum Hamming distance $d$. An $[n,k,d]$ code is called {\it distance-optimal} or {\it dimension-optimal} if there does not exist $[n,k,d+1]$ code or $[n,k+1,d]$ code, respectively. The Hamming weight of a codeword $\mathbf{c}=(c_0, c_1, \cdots, c_{n-1}) \in \mathcal{C}$ is the number of nonzero $c_i$ for $0\leq i\leq n-1$. Let $A_i$ denote the number of nonzero codewords with Hamming weight $i$ in $\mathcal{C}$. The {\it weight enumerator} of $\mathcal{C}$ is defined by $1+A_1x+A_2x^2+\cdots+A_nx^n$. The sequence $(1, A_1, \cdots, A_n)$ is called the {\it weight distribution} of $\mathcal{C}$. The weight distribution of a code
not only gives the error correcting ability of the code, but also allows the computation of the error probability
of error detection and correction~\cite{Klove2007}. Hence, the study of the weight distribution of a linear code is important
in both theory and applications. The reader can refer to \cite{Ding2015}-\cite{DDing2015}, \cite{Dinh2015} and the references therein.

 \vskip 6pt
It is well known that the minimum Hamming distances of linear codes play an important role in measuring error-correcting performance. Thus, how to find optimal linear codes with new lengths and minimum distances is one of the central topics in coding theory. In recent years, a series of work have been done.
Ding and Helleseth \cite{Ding2013} presented some optimal ternary cyclic codes of parameters $[3^m-1, 3m-2m-1, 4]$ according to the Sphere Packing bound by utilizing almost perfect nonlinear monomials and a number of other monomials over $\mathbb{F}_{3^m}$. Moreover, they proposed nine open problems about optimal ternary cyclic codes. Recently, three of those were solved~\cite{Han2019,Li2014,Li2015} by solving the certain equations
over finite fields and some new optimal ternary cyclic codes with parameters $[3^m-1, 3m-2m-1, 4]$ and $[3^m-1, 3m-2m-2,5]$ were obtained. Along this line, some other $p$-ary optimal linear codes have been studied~\cite{Fanarxiv, Fan2016, Xu2016, Zhou2019}. From these results, it is easy to see that most of the work in fining optimal linear codes focuses on constructing optimal ternary linear codes since analyzing the solutions of certain equations over $\mathbb{F}_{3}$ is easier than analyzing the solutions of certain equations over $\mathbb{F}_{p}$, where $p>3$. In this paper,  we obtain a lot of $p$-ary optimal linear codes for any odd prime $p$
by using the Pless power moment identities and the weight distributions of the duals of the linear codes.

\vskip 6pt
 To the best of our knowledge, subfield codes were first considered in \cite{Canteaut2000} and \cite{Carlet1998}, but the authors do not use the name ``subfield codes". The definition of subfield codes were first given by \cite[p.5117]{Cannon2013} and a Magma function for subfield codes is actual operated in the Magma package. Recently, following the Magma definition of subfield codes, Ding and Heng in \cite{Dingar} proved some basic results about the subfield codes of linear codes and determined the weight distributions of the subfield codes of ovoid codes.

 \vskip 6pt
 Let ${\rm PG}(2,\mathbb{F}_{p^m})$ denote the projective plane over the finite field $\mathbb{F}_{p^m}$.
 An {\it oval} in ${\rm PG}(2,\mathbb{F}_{p^m})$ is a set of ${p^m}+1$ points such that no three of them are collinear. A {\it conic} in ${\rm PG}(2,\mathbb{F}_{p^m})$ is a set of ${p^m}+1$ points that are zeros of a nondegenerate quadratic form in three variables. Obviously, a conic is an oval. Segre \cite{Segre1955} states that in ${\rm PG}(2,\mathbb{F}_{p^m})$ every oval is also a conic if $p$ is odd. Define a conic as follows:
 \begin{equation*}
 \begin{split}
 \mathcal{O}=\{(x^2,x,1): x \in \mathbb{F}_{p^m}\}\cup\{(1,0,0)\}.
 \end{split}
 \end{equation*}
 Assume that $\mathbb{F}_{p^m}=\{x_1,x_2,\cdots,x_{p^m}\}$ and $f(x)$ is a polynomial over $\mathbb{F}_{p^m}$. Let $\mathcal{ C}$ be a $[p^m+1,3]$ linear code with the generator matrix
 \begin{equation}\label{eq:matrixA'}
\begin{split}
 G_f  = \left(
\begin{array}{cccccc}
    f(x_1) & f(x_2) & \cdots & f(x_{p^m}) & 1  \\
     x_1 & x_2 & \cdots & x_{p^m} & 0  \\
     1 & 1 & \cdots & 1 & 0
\end{array}
\right).
\end{split}
\end{equation}
In \cite{Hengar}, Heng and Ding constructed a class of conic codes of length $p^m+1$ over $\mathbb{F}_{p^m}$ with generator matrix in (\ref{eq:matrixA'}) for $f(x)=x^2$,
and determined the weight distributions of the subfield codes of the conic codes.

 \vskip 6pt
Along the line of the work in~\cite{Hengar}, this paper constructs a class of linear codes of length $p^m+1$ over $\mathbb{F}_{p^m}$ with generator matrix
in (\ref{eq:matrixA'}) for $f(x)=x^{p^k+1}$, and
determines the weight distributions of the subfield codes of these linear codes, where the subfield code is defined as
\begin{equation}\label{code0}
\begin{split}
\mathcal{C}_k=\left\{\left(\left( {\rm Tr}_1^m\left(ax^{p^k+1}+bx\right)+c\right)_{x \in \mathbb{F}_{p^m}}, {\rm Tr}_1^m(a)\right) : \, a,b \in \mathbb{F}_{p^m}, c \in \mathbb{F}_p\right\}.
\end{split}
\end{equation}
Moreover, from the weight distribution of $\mathcal{C}_k$, we obtain the weight distribution of linear code

$$\mathcal{\bar{C}}_k=\left\{\left( {\rm Tr}_1^m\left(a x^{{p^k}+1}+bx\right)+c\right)_{x \in \mathbb{F}_{p^m}} : \, a,b \in \mathbb{F}_{p^m}, \,\,c \in \mathbb{F}_p\right\}.$$

\vskip 6pt
Hence, we generalize the results about the weight distributions of the subfield codes of conic codes in \cite{Hengar}. The parameters of these $p$-ary linear codes are new in some cases.  Some of the codes presented are optimal or almost optimal. Moreover, from the weight distributions of $\mathcal{C}_k$ and $\mathcal{\bar{C}}_k$ and the Pless power moment identities, we obtian that the duals of $\mathcal{C}_k$ and $\mathcal{\bar{C}}_k$ are optimal with respect to the Sphere Packing bound for many cases.

\vskip 6pt
The remainder of this paper is organized as follows. Section~$2$ recalls some preliminary results.
In Section~$3$ $($Section $4$$)$, the weight distribution of $\mathcal{C}_k$ and the dual of the parameters of $\mathcal{C}_k$  for the case $v_2(m)\leq v_2(k)$ $(v_2(m)> v_2(k))$ are determined, where $v_2(\cdot)$ denotes the 2-adic order function and $v_2(0)=\infty$. In Section $5$, from the weight distribution of $\mathcal{C}_k$ and the Sphere Packing bound, we obtain the weight distribution of $\mathcal{\bar{C}}_k$ and  the parameters of the dual of $\mathcal{\bar{C}}_k$. Section 6 concludes the paper.

\section{Preliminaries}

Let $\mathbb{F}_{p^m}$ be a finite field and $\mathbb{F}_{p^m}^* = \mathbb{F}_{p^m} \setminus \{0\}$. Let $\psi$ be a multiplicative character of $\mathbb{F}_{p^m}^*$ and $\chi$ be a canonical additive character of $\mathbb{F}_{p^m}$.
The {\it Gaussian sum} $G(\psi, \chi)$ is defined by
\[ G(\psi, \chi) = \sum_{x\in \mathbb{F}_{p^m}^*} \psi(x)\chi(x) .\]
 It is very difficult to determine explicit values of Gaussian sums in general. Until now, Gaussian sums are known only for a few special cases. The  quadratic Gaussian sums are known and given in the following lemma.
\begin{lemma}\cite[Theorem 5.15]{Lidl1983}\label{lemGxn}
Let $\mathbb{F}_{p^m}$ be a finite field. Then
\[G(\eta,\chi)=\left\{ \begin{array}{lcl}
           (-1)^{m-1}p^{\frac{m}{2}}, & {\rm if}\, \,\, p\equiv 1 \pmod 4,\\
           (-1)^{m-1}i^mp^{\frac{m}{2}}, & {\rm if}\, \,\, p\equiv 3 \pmod 4, \end{array}  \right.\]
\end{lemma}
where $\eta$ is the quadratic multiplicative character of $\mathbb{F}_{p^m}$.

\vskip 6pt
Let $\mathbb{F}_q$ denote the finite field with $q$ elements, where $q$ is an odd prime power. By identifying the finite field $\mathbb{F}_{q^s}$ with an $s$-dimensional vector space
$\mathbb{F}_{q}^s$ over $\mathbb{F}_{q}$, a function~$f$ from $\mathbb{F}_{q^s}$ to $\mathbb{F}_q$ can be viewed as an $s$-variable polynomial over $\mathbb{F}_{q}$.
The function $f(x)$ is called a quadratic form
if it is a homogenous polynomial of degree two as follows:
\[f(x_1, x_2, \cdots, x_s) = \sum_{1\leq i\leq j\leq s} a_{ij} x_ix_j , \,\, \, a_{ij}\in \mathbb{F}_{q},\]
where we fix  a  basis of $\mathbb{F}_q^s$ over $\mathbb{F}_q$ and identify $x\in \mathbb{F}_{q^s}$ with a vector $(x_1, x_2, \cdots, x_s)\in \mathbb{F}_q^s$.
The rank of the quadratic form $f(x)$ is defined as the codimension of $\mathbb{F}_q$-vector space
\[ V = \{ x\in \mathbb{F}_q^s\, \, |\,\, f(x+z)-f(x)-f(z) = 0, \,\, \mbox{for all}\,\, z\in \mathbb{F}_{q}^s \} ,\]
which is denoted by rank$(f)$. Then $|V| = q^{s-{\rm rank}(f)}$.
\vskip 6pt
The following lemma gives a general result on an exponential sum of a quadratic function from $\mathbb{F}_{p^m}$ to $\mathbb{F}_p$.

\begin{lemma}\label{lem:quadraticsum} (\cite[Lemma 2.1]{Liu2018})
Let $Q(x)$ be a quadratic function from $\mathbb{F}_{p^m}$ to $\mathbb{F}_p$ with rank $r(r\neq0)$, $\zeta_p$ be a $p$-th primitive root of unity and
$\eta$ be the quadratic multiplicative character of $\mathbb{F}_{p^m}$. Then
 for any $z\in \mathbb{F}_{p}^*$,
\begin{equation*}
\sum_{x\in \mathbb{F}_{p^m}} \zeta_p^{z Q(x)}=\eta^r(z)\sum_{x\in \mathbb{F}_{p^m}} \zeta_p^{ Q(x)}.
\end{equation*}
\end{lemma}

The following is a well known result about quadratic forms.

\begin{lemma}\label{cor:quadraticsum}\cite[Corollary 7.6]{Draper2007}
Let $v_2(\cdot)$ denote the 2-adic order function and $v_2(0)=\infty$. Let $a \in \mathbb{F}_{p^m}^*$ and $k$ be a nonnegative integer. Assume that $Q(x)={\rm Tr}_1^m\left(ax^{p^k+1}\right)$.
\begin{description}
\item{(i)} If $v_2(m)\leq v_2(k)$,
\begin{equation*}
\begin{split}
\sum_{x\in \mathbb{F}_{p^m}} \zeta_p^{Q(x)}=\eta(a)(-1)^{m-1}i^{\frac{(p-1)^2m}{2}}p^{\frac{m}{2}}.
\end{split}
\end{equation*}
\item{(ii)} If $v_2(m)=v_2(k)+1$,
\begin{equation*}
\begin{split}
\sum_{x\in \mathbb{F}_{p^m}} \zeta_p^{Q(x)}=
\begin{cases}
p^{\frac{m+\gcd(2k,m)}{2}}, & \text{if\,\,\,\, $a^{\frac{(p^k-1)(p^m-1)}{p^{\gcd(2k,m)}-1}}=-1$,} \\
-p^{\frac{m}{2}}, & \text{otherwise}.
\end{cases}
\end{split}
\end{equation*}
\item{(iii)} If $v_2(m)>v_2(k)+1$,
\begin{equation*}
\begin{split}
\sum_{x\in \mathbb{F}_{p^m}} \zeta_p^{Q(x)}=
\begin{cases}
-p^{\frac{m+\gcd(2k,m)}{2}}, & \text{if\,\,\,\, $a^{\frac{(p^k-1)(p^m-1)}{p^{\gcd(2k,m)}-1}}=1$,} \\
p^{\frac{m}{2}}, & \text{otherwise}.
\end{cases}
\end{split}
\end{equation*}
\end{description}
\end{lemma}

Next, we recall some known results about the subfield codes, which were introduced in \cite{Dingar}. Let $\mathcal{C}$ be an $[n,k]$ code over $\mathbb{F}_{q^m}$ with the generator matrix

 \begin{equation*}
\begin{split}
G= \left(
\begin{array}{cccc}
    g_{11} & g_{12} & \cdots & g_{1n} \\
     g_{21} & g_{22} & \cdots & g_{2n} \\
     \vdots & \vdots & \ddots & \vdots\\
     g_{k1} & g_{k2} & \cdots & g_{kn}
\end{array}
\right).
\end{split}
\end{equation*}
Assume that $\{\alpha_1, \alpha_2, \cdots, \alpha_m\}$ is a basis of $\mathbb{F}_{q^m}$ over $\mathbb{F}_q$. Ding and Heng in \cite{Dingar} constructed a new $[n,k']$ code $\mathcal{C}^{(q)}$ over $\mathbb{F}_q$ with
 the generator matrix
 \begin{equation*}
\begin{split}
G^{(q)}= \left(
 \begin{array}{cccc}
   G_1^{(q)}  \\
     G_2^{(q)}  \\
     \vdots \\
     G_k^{(q)}
\end{array}
\right),
\end{split}
\end{equation*}
where
 \begin{equation*}
\begin{split}
\ G^{(q)}_i = \left(
\begin{array}{cccc}
    {\rm Tr}_1^m(g_{i1}\alpha_1) & {\rm Tr}_1^m(g_{i2}\alpha_1) & \cdots & {\rm Tr}_1^m(g_{in}\alpha_1) \\
     {\rm Tr}_1^m(g_{i1}\alpha_2) & {\rm Tr}_1^m(g_{i2}\alpha_2) & \cdots & {\rm Tr}_1^m(g_{in}\alpha_2) \\
     \vdots & \vdots & \ddots & \vdots\\
    {\rm Tr}_1^m(g_{i1}\alpha_m) & {\rm Tr}_1^m(g_{i2}\alpha_m) & \cdots & {\rm Tr}_1^m(g_{in}\alpha_m) \\
\end{array}
\right).
\end{split}
\end{equation*}
They called $\mathcal{C}^{(q)}$ the {\it subfield code} of $\mathcal{C}$ and gave the trace representation of $\mathcal{C}^{(q)}$ as follows.
\begin{lemma}\cite[Thoerem 2.5] {Dingar}\label{subfieldcode}
Let $\mathcal{C}$ be an $[n,k]$ code over $\mathbb{F}_{q^m}$. Let $G=[g_{i,j}]_{1\leq i \leq k, 1\leq j \leq n}$ be a generator matrix of $\mathcal{C}$. Then the trace representation of the subfield code $\mathcal{C}^{(q)}$ is given by
$$\mathcal{C}^{(q)}=\left\{\left( {\rm Tr}_1^m\left(\sum_{i=1}^ka_ig_{i1}\right),\cdots, {\rm Tr}_1^m\left(\sum_{i=1}^ka_ig_{i1}\right)\right)\,:\, a_1,\cdots, a_k \in \mathbb{F}_{q^m}\right\}.$$
\end{lemma}

Let $q=p$ be an odd prime and the generator matrix of $\mathcal{C}$ be defined in (\ref{eq:matrixA'}). From Lemma \ref{subfieldcode} we have the trace representation of the subfield code of $\mathcal{C}$ as follows:
\begin{equation*}
\begin{split}
\mathcal{C}_f=\left\{\left(\left( {\rm Tr}_1^m(af(x)+bx)+c\right)_{x \in \mathbb{F}_{p^m}}, {\rm Tr}_1^m(a)\right)\, : \, a,b \in \mathbb{F}_{p^m}, c \in \mathbb{F}_p\right\}.
\end{split}
\end{equation*}

When $a=0$, the Hamming weights and their corresponding frequencies of the codewords of the form $\mathbf{c}(0,b,c)$ can be easily determined as follows.
\vskip 6pt
If $a=b=0$, then
\begin{equation*}
\begin{split}
\mathbf{c}(0,0,c)=(\underbrace{c,c, \cdots, c}_{\small{p^m}},0),
\end{split}
\end{equation*}
i.e.,
\begin{equation}\label{111}
\begin{split}
\wt(\mathbf{c}(0,0,c))=
\begin{cases}
0, & \text{occur\,\,\,\, $1$\,\,\,\,\,\,\, time,} \\
p^m, & \text{occur $p-1$ times}.
\end{cases}
\end{split}
\end{equation}
If $a=0$, $b\neq 0$, then
\begin{equation*}
\begin{split}
\mathbf{c}(0,b,c)=\left(\left({\rm Tr}_1^m(bx)+c\right)_{x \in \mathbb{F}_{p^m}}, 0\right).
\end{split}
\end{equation*}
It is easy for us to get that
\begin{equation}\label{112}
\begin{split}
\wt(\mathbf{c}(0,b,c))=
p^{m-1}(p-1)\,\,\, & \text{occur} \,\, p(p^m-1) \,\, \text{times}.
\end{split}
\end{equation}

When $a\neq0$, to determine the Hamming weights and their corresponding frequencies of $\mathbf{c}(a,b,c)$ in $\mathcal{C}_f$ for $(a,b,c)$ running through
$(\mathbb{F}_{p^m}^*, \mathbb{F}_{p^m}, \mathbb{F}_p)$ is a very hard problem for the general polynomial $f(x)$. In the following sections, we always assume that $f(x)=x^{p^k+1}$,  where $k$ is a nonnegative integer.
\vskip 6pt

 In order to obtain the parameters of the dual codes of the discussed subfield codes, we need the Pless power moment
identities on linear codes. Let $\mathcal{C}$ be a $p$-ary $[n, k]$ code over $\mathbb{F}_p$, and denote its dual by $\mathcal{C}^{\perp}$. Let
 $A_i$ and $A^{\perp}_i$ be the number of codewords of weight $i$ in $\mathcal{C}$ and $\mathcal{C}^{\perp}$, respectively.
 The first five Pless power moment identities are as follows (\cite{MacWilliam1997}, p. 131):
\begin{equation*}
\begin{split}
&\sum_{i=0}^nA_i=p^k;\\
&\sum_{i=0}^niA_i=p^{k-1}(pn-n-A_1^{\perp});\\
&\sum_{i=0}^ni^2A_i=p^{k-2}[(p-1)n(pn-n+1)-(2pn-p-2n+2)A_1^{\perp}+2A_2^{\perp}];\\
&\sum_{i=0}^ni^3A_i=p^{k-3}[(p-1)n(p^2n^2-2pn^2+3pn-p+n^2-3n+2)-(3p^2n^2-3p^2n-6pn^2+12pn\\
&\hskip 48pt+p^2-6p+3n^2-9n+6)A_1^{\perp}+6(pn-p-n+2)A_2^{\perp}-6A_3^{\perp}];\\
&\sum_{i=0}^ni^4A_i=p^{k-4}[(p-1)n(p^3n^3-3p^2n^3+6p^2n^2-4p^2n+p^2+3pn^3-12pn^2+15pn-6p-n^3+6n^2-11n+6)\\
&\hskip 48pt-(4p^3n^3-6p^3n^2+4p^3n-p^3-12p^2n^3+36p^2n^2-38p^2n+14p^2+12pn^3-54pn^2+78pn-36p\\
&\hskip 48pt-4n^3+24n^2-44n+24)A_1^{\perp}+(12p^2n^2-24p^2n+14p^2-24pn^2+84pn-72p+12n^2-60n+72)A_2^{\perp}\\
&\hskip 48pt-(24pn-36p-24n+72)A_3^{\perp}+24A_4^{\perp}].
\end{split}
\end{equation*}

 The following two lemmas on the bound of linear codes are well-known.

\begin{lemma}(Sphere Packing bound)
Let $\mathcal{C}$ be a $p$-ary $[n, k,d]$ code. Then
$$p^n\geq p^k\sum_{i=0}^{\lfloor \frac{d-1}{2} \rfloor}\left(
\begin{array}{cccc}
   n  \\
     i  \\
\end{array}
\right)(p-1)^i.$$
\end{lemma}

\begin{lemma}\cite{Rouayheb2007}\label{bound2}
Let $q$ be an odd prime power and $A_q(n,d)$ be the maximum number of codewords of a $q$-ary code with length $n$ and Hamming distance at least $d$. If $q\geq 3$, $t=n-d+1$ and $r=\lfloor min\{\frac{n-t}{2}, \frac{t-1}{q-2}\}\rfloor$, then
\begin{equation*}
\begin{split}
A_q(n,d)\leq \frac{q^{t+2r}}{\sum_{i=0}^r\left(\begin{array}{cccc}
   t+2r  \\
     i  \\
\end{array}
\right)(q-1)^i}.
\end{split}
\end{equation*}
\end{lemma}


\section{The weight distribution of $\mathcal{C}_k$ for $v_2(m)\leq v_2(k)$}
Through out this section, let $\eta$ and $\eta_0$ denote the quadratic multiplicative character of $\mathbb{F}_{p^m}$ and $\mathbb{F}_p$, respectively. Our main task is to determine the weight distribution of $\mathcal{C}_k$ for the case $v_2(m)\leq v_2(k)$ and give the parameters of the dual of $\mathcal{C}_k$, where $\mathcal{C}_k$ is defined in (\ref{code0}). We start with the following lemma, which is given by \cite{Coulter2002} and \cite{Lidl1983}.

\begin{lemma}\cite{Coulter2002,Lidl1983}\label{eqSab}
Let $v_2(\cdot)$ denote the 2-adic order function and $v_2(0)=\infty$. Let $a\in \mathbb{F}_{p^m}^*$, $b \in \mathbb{F}_{p^m}$, and $k,m$ be two integers such that $v_2(m)\leq v_2(k)$. Let
\begin{equation*}
\begin{split}
S(a,b)=\sum_{x \in \mathbb{F}_{p^m}}\zeta_p^{{\rm Tr}_1^m\left(ax^{p^k+1}+bx\right)},
\end{split}
\end{equation*} and $x_b$ be the unique solution of the equation $a^{p^k}x^{p^{2k}}+ax+b^{p^k}=0$. Then
\[S(a,b)=\left\{ \begin{array}{lcl}
           (-1)^{m-1}p^{\frac{m}{2}}\eta(-a)\zeta_p^{{\rm Tr}_1^m\left(-ax_b^{p^k+1}\right)}, & {\rm if}\, \,\, p\equiv 1 \pmod 4,\\
           (-1)^{m-1}p^{\frac{m}{2}}i^{3m}\eta(-a)\zeta_p^{{\rm Tr}_1^m\left(-ax_b^{p^k+1}\right)}, & {\rm if}\, \,\, p\equiv 3 \pmod 4, \end{array}  \right.\]
where $\zeta_p$ is a $p$-th primitive root of unity.
\end{lemma}

\begin{remark}
For any $a \in \mathbb{F}_{p^m}^*$, it is clear that $a^{p^k}x^{p^{2k}}+ax=0$ is a linearized permutation polynomial over $\mathbb{F}_{p^m}$ for the case $v_2(m)\leq v_2(k)$. Hence, the equation $a^{p^k}x^{p^{2k}}+ax+b^{p^k}=0$ has a unique solution.
\end{remark}

Now, we determine the possible values of $\wt(\mathbf{c}(a,b,c))$ for $(a,b,c)$ running through $(\mathbb{F}_{p^m}^*,\mathbb{F}_{p^m}, \mathbb{F}_p)$.

\begin{proposition}\label{pro:ww}
Let $a\in \mathbb{F}_{p^m}^*$, $b \in \mathbb{F}_{p^m}$ and $c \in \mathbb{F}_p$. Let $x_b$ be the unique solution of the equation $a^{p^k}x^{p^{2k}}+ax+b^{p^k}=0$. Assume that $v_2(m)\leq v_2(k)$ and $c_0=c-{\rm Tr}_1^m\left(ax_b^{p^k+1}\right)$. If  $v_2(m)=0$, then the possible distinct values of $\wt(\mathbf{c}(a,b,c))$ are
\begin{equation*}
\begin{split}
\wt(\mathbf{c}(a,b,c))=\begin{cases}
(p-1)p^{m-1}, & \text{if \,\,${\rm Tr}_1^m(a)=0$ and $c_0=0$}, \\
(p-1)p^{m-1}+1, & \text{if \,\,${\rm Tr}_1^m(a)\neq 0$ and $c_0= 0$}, \\
(p-1)p^{m-1}-p^{\frac{m-1}{2}}, & \text{if \,\,${\rm Tr}_1^m(a)=0$ and $\eta(ac_0)=1$,} \\
(p-1)p^{m-1}-p^{\frac{m-1}{2}}+1, & \text{if \,\,${\rm Tr}_1^m(a)\neq 0$ and $\eta(ac_0)=1$,} \\
(p-1)p^{m-1}+p^{\frac{m-1}{2}}, & \text{if \,\,${\rm Tr}_1^m(a)=0$ and $\eta(ac_0)=-1$,} \\
(p-1)p^{m-1}+p^{\frac{m-1}{2}}+1, & \text{if \,\,${\rm Tr}_1^m(a)\neq 0$ and $\eta(ac_0)=-1$.} \\
\end{cases}
\end{split}
\end{equation*}
If $0<v_2(m)\leq v_2(k)$, then the possible distinct values of $\wt(\mathbf{c}(a,b,c))$ are
\begin{equation*}
\begin{split}
\wt(\mathbf{c}(a,b,c))=\begin{cases}
(p-1)p^{m-1}-\varepsilon_0(p-1)p^{\frac{m-2}{2}}, & \text{if \,\,${\rm Tr}_1^m(a)=0$, $\eta(a)=-1$ and $c_0=0$,} \\
(p-1)p^{m-1}-\varepsilon_0(p-1)p^{\frac{m-2}{2}}+1, & \text{if \,\,${\rm Tr}_1^m(a)\neq0$, $\eta(a)=-1$ and $c_0=0$,} \\
(p-1)p^{m-1}-\varepsilon_0p^{\frac{m-2}{2}}, & \text{if \,\,${\rm Tr}_1^m(a)=0$, $\eta(a)=1$ and $c_0\neq0$,} \\
(p-1)p^{m-1}-\varepsilon_0p^{\frac{m-2}{2}}+1, &  \text{if \,\,${\rm Tr}_1^m(a)\neq0$, $\eta(a)=1$ and $c_0\neq0$,} \\
(p-1)p^{m-1}+\varepsilon_0(p-1)p^{\frac{m-2}{2}}, &\text{if \,\,${\rm Tr}_1^m(a)=0$, $\eta(a)=1$ and $c_0=0$,} \\
(p-1)p^{m-1}+\varepsilon_0(p-1)p^{\frac{m-2}{2}}+1, & \text{if \,\,${\rm Tr}_1^m(a)\neq0$, $\eta(a)=1$ and $c_0=0$,} \\
(p-1)p^{m-1}+\varepsilon_0p^{\frac{m-2}{2}}, & \text{if \,\,${\rm Tr}_1^m(a)=0$, $\eta(a)=-1$ and $c_0\neq0$,} \\
(p-1)p^{m-1}+\varepsilon_0p^{\frac{m-2}{2}}+1, &  \text{if \,\,${\rm Tr}_1^m(a)\neq0$, $\eta(a)=-1$ and $c_0\neq0$,} \\
\end{cases}
\end{split}
\end{equation*}
where $\varepsilon_0=(-1)^{\frac{(p-1)m}{4}}$.
\end{proposition}
{\it Proof}.
 The weight of the codeword $\mathbf{c}(a,b,c)$ is related to the exponential sum $S(a,b)$. According to Lemma~\ref{eqSab}, we only discuss the case $p \equiv 1\pmod 4$, and the case $p \equiv 3\pmod 4$ can be shown by the similar way. By the definition of $\mathcal{C}_k$, the last entry of codeword $\mathbf{c}(a,b,c)$ is $0$ (or not $0$) if ${\rm Tr}_1^m(a)=0$ ( or ${\rm Tr}_1^m(a)\neq 0$). So, we prove the theorem
 according to the following four cases.
\begin{description}
\item{\bf Case 1}: $v_2(m)=0$ and ${\rm Tr}_1^m(a)=0$.  By Lemma \ref{eqSab} we have

\begin{equation}\label{eq:cabc1}
\begin{split}
\wt(\mathbf{c}(a,b,c))&=p^m-\frac{1}{p}\sum_{z \in \mathbb{F}_p}\sum_{x \in \mathbb{F}_{p^m}}\zeta_p^{z\left({\rm Tr}_1^m\left(ax^{p^k+1}+bx\right)+c\right)}\\
&=(p-1)p^{m-1}-\frac{1}{p}\sum_{z \in \mathbb{F}_p^*}\sum_{x \in \mathbb{F}_{p^m}}\zeta_p^{z\left({\rm Tr}_1^m\left(ax^{p^k+1}+bx\right)+c\right)}\\
&=(p-1)p^{m-1}-\frac{1}{p}\sum_{z \in \mathbb{F}_p^*}\zeta_p^{zc}\sum_{x \in \mathbb{F}_{p^m}}\zeta_p^{{\rm Tr}_1^m\left(zax^{p^k+1}+zbx\right)}\\
&=(p-1)p^{m-1}-p^{\frac{m-2}{2}}\sum_{z \in \mathbb{F}_p^*}\eta(-za)\zeta_p^{z\left(c-{\rm Tr}_1^m\left(ax_b^{p^k+1}\right)\right)}.
\end{split}
\end{equation}

If $c={\rm Tr}_1^m\left(ax_b^{p^k+1}\right)$, then $\wt(\mathbf{c}(a,b,c))=(p-1)p^{m-1}$.

If $c\neq{\rm Tr}_1^m\left(ax_b^{p^k+1}\right)$, recalling that $c_0=c-{\rm Tr}_1^m\left(ax_b^{p^k+1}\right)$, by Lemma~\ref{lemGxn} and (\ref{eq:cabc1}) we have
\begin{equation*}
\begin{split}
\wt(\mathbf{c}(a,b,c))&=(p-1)p^{m-1}-p^{\frac{m-2}{2}}\sum_{z \in \mathbb{F}_p^*}\eta(-za)\zeta_p^{zc_0}\\
&=(p-1)p^{m-1}-p^{\frac{m-2}{2}}\sum_{z \in \mathbb{F}_p^*}\eta(-za)\eta^2 (c_0)\zeta_p^{zc_0}\\
&=(p-1)p^{m-1}-p^{\frac{m-2}{2}}\eta(-ac_0)\sum_{z \in \mathbb{F}_p^*}\eta(zc_0)\zeta_p^{zc_0}\\
&=(p-1)p^{m-1}-p^{\frac{m-2}{2}}\eta(ac_0)\sum_{z \in \mathbb{F}_p^*}\eta_0(zc_0)\zeta_p^{zc_0}\\
&=(p-1)p^{m-1}-p^{\frac{m-2}{2}}\eta(ac_0)G(\eta_0,\chi_0)\\
&=\begin{cases}
(p-1)p^{m-1}-p^{\frac{m-1}{2}}, & \text{if $\eta(ac_0)=1$,} \\
(p-1)p^{m-1}+p^{\frac{m-1}{2}}, & \text{if $\eta(ac_0)=-1$,} \\
\end{cases}
\end{split}
\end{equation*}
 where $\chi_0$ is a canonical additive character of $\mathbb{F}_{p}$.
In the fourth equality, we used the fact that $\eta_0(zc_0)=\eta(zc_0)$ and $\eta(-1)=1$ since $m$ is odd and $p\equiv1 \pmod 4$.

\item{\bf Case 2}: $v_2(m)=0$ and ${\rm Tr}_1^m(a)\neq0$. Then $\wt(\mathbf{c}(a,b,c))$ can be expressed as
\begin{equation*}
\begin{split}
\wt(\mathbf{c}(a,b,c))=p^m+1-\frac{1}{p}\sum_{z \in \mathbb{F}_p}\sum_{x \in \mathbb{F}_{p^m}}\zeta_p^{z\left({\rm Tr}_1^m\left(ax^{p^k+1}+bx\right)+c\right)}.
\end{split}
\end{equation*}
Using the same way as in Case $1$, we have
\begin{equation*}
\begin{split}
\wt(\mathbf{c}(a,b,c))
&=\begin{cases}
(p-1)p^{m-1}+1, & \text{if $c_0=0$,} \\
(p-1)p^{m-1}-p^{\frac{m-1}{2}}+1, & \text{if $\eta(ac_0)=1$,} \\
(p-1)p^{m-1}+p^{\frac{m-1}{2}}+1, & \text{if $\eta(ac_0)=-1$.} \\
\end{cases}
\end{split}
\end{equation*}

\item{\bf Case 3}: $0<v_2(m)\leq v_2(k)$ and ${\rm Tr}_1^m(a)=0$.  By Lemma \ref{eqSab} and (\ref{eq:cabc1}) we have
\begin{equation*}
\begin{split}
\wt(\mathbf{c}(a,b,c))&=(p-1)p^{m-1}+p^{\frac{m-2}{2}}\sum_{z \in \mathbb{F}_p^*}\eta(-za)\zeta_p^{z\left(c-{\rm Tr}_1^m\left(ax_b^{p^k+1}\right)\right)}\\
&=(p-1)p^{m-1}+p^{\frac{m-2}{2}}\eta(a)\sum_{z \in \mathbb{F}_p^*}\zeta_p^{z\left(c-{\rm Tr}_1^m\left(ax_b^{p^k+1}\right)\right)}\\
&=\begin{cases}
(p-1)p^{m-1}+(p-1)p^{\frac{m-2}{2}}, & \text{if $\eta(a)=1$ and $c_0 =0$,} \\
(p-1)p^{m-1}-(p-1)p^{\frac{m-2}{2}}, & \text{if $\eta(a)=-1$ and $c_0 =0$,} \\
(p-1)p^{m-1}-p^{\frac{m-2}{2}}, & \text{if $\eta(a)=1$ and $c_0 \neq0$,} \\
(p-1)p^{m-1}+p^{\frac{m-2}{2}}, & \text{if $\eta(a)=-1$ and $c_0 \neq0$.} \\
\end{cases}
\end{split}
\end{equation*}
In the second equality, we used the fact that $\eta(-z)=1$ for $z \in \mathbb{F}_p^*$ since $m$ is even.

\item{\bf Case 4}: $0<v_2(m)\leq v_2(k)$ and ${\rm Tr}_1^m(a)\neq0$. Using the same way as in Case $3$,  we have
\begin{equation*}
\begin{split}
\wt(\mathbf{c}(a,b,c))
=\begin{cases}
(p-1)p^{m-1}+(p-1)p^{\frac{m-2}{2}}+1, & \text{if $\eta(a)=1$ and $c_0 =0$,} \\
(p-1)p^{m-1}-(p-1)p^{\frac{m-2}{2}}+1, & \text{if $\eta(a)=-1$ and $c_0 =0$,} \\
(p-1)p^{m-1}-p^{\frac{m-1}{2}}+1, & \text{if $\eta(a)=1$ and $c_0 \neq0$,} \\
(p-1)p^{m-1}+p^{\frac{m-1}{2}}+1, & \text{if $\eta(a)=-1$ and $c_0 \neq0$.} \\
\end{cases}
\end{split}
\end{equation*}

From Case $1$, Case $2$, Case $3$ and Case $4$,
the result follows. $\square$
\end{description}

In the following, we determine the weight distribution of $\mathcal{C}_k$. To this end, we first determine the lower bound of the minimum Hamming distance of $\mathcal{C}_k^{\perp}$.

\begin{lemma}
Let $\mathcal{C}_k$ be the linear code defined in (\ref{code0}), then the minimum Hamming distance of $\mathcal{C}_k^{\perp}$ is greater than or equal to $3$.
\end{lemma}
{\it Proof.}  We only need to prove that there does not exist a codeword $\mathbf{c}^{\perp}\in\mathcal{C}_k^{\perp}$ with  the Hamming weight $1$ and $2$. There are following cases.
\begin{description}
\item{\bf Case 1}: Assume that $a=b=0$ and $c =1$ in (\ref{code0}), then
$\mathbf{c}=(\mathbf{1},0) \in \mathcal{C}_k,$ where $\mathbf{1}$ is a vector of length $p^m$ with all entries being $1$.
If there exists a codeword $\mathbf{c}^{\perp}\in \mathcal{C}_k^{\perp}$ with Hamming weight $1$, then $\mathbf{c}^{\perp}$ must have the form
$(\mathbf{0},\alpha) \in \mathcal{C}_k^{\perp},$ where $\mathbf{0}$ is a vector of length $p^m$ with all entries being $0$ and $\alpha \in \mathbb{F}_p^*$.
It is obvious that for any ${\rm Tr}_1^m(a)\neq 0$, there is a codeword  $(\mathbf{u},{\rm Tr}_1^m(a)) \in \mathcal{C}_k$ such that $(\mathbf{u},{\rm Tr}_1^m(a))(\mathbf{0},\alpha)\neq 0$, where $\mathbf{u}$ is a vector of length $p^m$. Hence, there is not a codeword in $\mathcal{C}_k^{\perp}$ with Hamming weight $1$.
\item{\bf Case 2}:  Assume that there is a codeword $\mathbf{c}^{\perp} \in \mathcal{C}^{\perp}$ with ${\rm wt_H}(\mathbf{c}^{\perp})=2$. Obviously, $\mathbf{c}^{\perp}$ must have the form $(\mathbf{0},\alpha,\mathbf{0},-\alpha,\mathbf{0})$ since $(\mathbf{1},0) \in \mathcal{C}$, where $\alpha \in \mathbb{F}_p^*$. So, let $c=0$, for any $a,b \in \mathbb{F}_{p^m}$, there exist two different elements $\beta,\gamma \in \mathbb{F}_{p^m}$ such that
    $${\rm Tr}_1^m\left(a\beta^{p^k+1}+b\beta\right)=-{\rm Tr}_1^m\left(a\gamma^{p^k+1}+b\gamma\right).$$
 This means that
    \begin{equation}\label{gamma}
    \begin{split}
    {\rm Tr}_1^m\left(a\left(\beta^{p^k+1}+\gamma^{p^k+1}\right)+b(\beta+\gamma)\right)=0
    \end{split}
    \end{equation}
    for any $a,b \in \mathbb{F}_{p^m}.$
It is easy to see that (\ref{gamma}) holds if and only if
\begin{equation*}
\begin{split}
\begin{cases}
\beta^{p^k+1}+\gamma^{p^k+1}=0, \\
\beta+\gamma=0,
\end{cases}
\end{split}
\end{equation*}
which is impossible.
\end{description}
Combining with Case $1$ and Case $2$, the result follows. $\square$

\vskip 6pt
The following result \cite{Hengar} is useful for us to determine the weight distribution of $\mathcal{C}_k$.
\begin{lemma}\cite[Lemma 14]{Hengar}\label{lemeven}
 Let $\mathbb{F}_{p^m}$ be the finite field of $p^m$ elements, where $p$ is an odd prime. Let $\eta$ be the quadratic multiplicative character of $\mathbb{F}_{p^m}^*$. If $m$ is even, then
\begin{equation*}
\begin{split}
\left|\left\{a \in \mathbb{F}_{p^m}^*\,|\, \eta(a)=\pm 1\,\, and\,\, {\rm Tr}_1^m(a)=0\right\}\right|=
\frac{p^{m-1}-1\mp (p-1)p^{\frac{m-2}{2}}(-1)^{\frac{(p-1)m}{4}}}{2},
\end{split}
\end{equation*}
and
\begin{equation*}
\begin{split}
\left|\{a \in \mathbb{F}_{p^m}^*\,|\, \eta(a)=\pm1\,\, and\,\, {\rm Tr}_1^m(a)\neq0\}\right|=
\frac{(p-1)\left(p^{m-1}\pm p^{\frac{m-2}{2}}(-1)^{{\frac{(p-1)m}{4}}}\right)}{2}.
\end{split}
\end{equation*}
\end{lemma}

With above preparations, we now give the weight distribution of $\mathcal{C}_k$ for the case $v_2(m)\leq v_2(k)$.
\begin{theorem}\label{theorem1}
 Let $k,m$ be two integers such that $v_2(m)\leq v_2(k)$. Let $\mathcal{C}_k$ be the linear code defined in (\ref{code0}).
 \begin{description}
 \item{(i)} If $v_2(m)=0$, then $\mathcal{C}_k$ is a $\left[p^m+1,2m+1,(p-1)p^{m-1}-p^{\frac{m-1}{2}}\right]$ code with
the weight distribution in Table~\ref{Table1}. Its dual has parameters $[p^m+1,p^m-2m,4]$, which is optimal with respect to the Sphere Packing bound for $p\geq 5$.
\begin{table}[h]
{\caption{\rm   The weight distribution of $\mathcal{C}_k$ for $v_2(m)=0$   }\label{Table1}
\begin{center}
\begin{tabular}{cccc}\hline
     Weight & Multiplicity \\\hline
  $0$ & $1$  \\
 $p^m$ & $p-1$ \\
  $(p-1)p^{m-1}$ & $p(p^m-1)+p^m(p^{m-1}-1)$ \\
  $(p-1)p^{m-1}+1$ & $p^m(p^m-p^{m-1})$ \\
 $(p-1)p^{m-1}\pm p^{\frac{m-1}{2}}$ & $p^m(p^{m-1}-1)(p-1)/2$ \\
$(p-1)p^{m-1}\pm p^{\frac{m-1}{2}}+1$ & $p^{2m-1}(p-1)^2/2$ \\   \hline
\end{tabular}
\end{center}}
\end{table}
\item{(ii)} If $0<v_2(m)\leq v_2(k)$, then $\mathcal{C}_k$ is a $\left[p^m+1,2m+1,(p-1)\left(p^{m-1}-p^{\frac{m-2}{2}}\right)\right]$ code with
the weight distribution in Table~\ref{Table2}. Its dual has parameters $[p^m+1,p^m-2m,4]$, which is optimal with respect to the Sphere Packing bound for $p\geq 5$.

\begin{table}[h]
{\caption{\rm   The weight distribution of $\mathcal{C}_k$ for $0<v_2(m)\leq v_2(k)$   }\label{Table2}
\begin{center}
\begin{tabular}{cccc}\hline
     Weight & Multiplicity \\\hline
  $0$ & $1$  \\
 $p^m$ & $p-1$ \\
  $(p-1)p^{m-1}$ & $p(p^m-1)$ \\
  $(p-1)\left(p^{m-1}+\varepsilon_0p^{\frac{m-2}{2}}\right)$ & $p^m\left(p^{m-1}-1-\varepsilon_0(p-1)p^{\frac{m-2}{2}}\right)/2$ \\
  $(p-1)\left(p^{m-1}+\varepsilon_0p^{\frac{m-2}{2}}\right)+1$ & $p^m(p-1)\left(p^{m-1}+\varepsilon_0p^{\frac{m-2}{2}}\right)/{2}$ \\
  $(p-1)\left(p^{m-1}-\varepsilon_0p^{\frac{m-2}{2}}\right)$ & $p^m\left(p^{m-1}-1+\varepsilon_0(p-1)p^{\frac{m-2}{2}}\right)/2$ \\
    $(p-1)\left(p^{m-1}-\varepsilon_0p^{\frac{m-2}{2}}\right)+1$ & $p^m(p-1)\left(p^{m-1}-\varepsilon_0p^{\frac{m-2}{2}}\right)/2$ \\
  $(p-1)p^{m-1}-\varepsilon_0p^{\frac{m-2}{2}}$ & $p^m(p-1)\left(p^{m-1}-1-\varepsilon_0(p-1)p^{\frac{m-2}{2}}\right)/2$ \\
  $(p-1)p^{m-1}-\varepsilon_0p^{\frac{m-2}{2}}+1$ & $p^m(p-1)^2\left(p^{m-1}+\varepsilon_0p^{\frac{m-2}{2}}\right)/2$ \\
  $(p-1)p^{m-1}+\varepsilon_0p^{\frac{m-2}{2}}$ & $p^m(p-1)\left(p^{m-1}-1+\varepsilon_0(p-1)p^{\frac{m-2}{2}}\right)/2$ \\
    $(p-1)p^{m-1}+\varepsilon_0p^{\frac{m-2}{2}}+1$ & $p^m(p-1)^2\left(p^{m-1}-\varepsilon_0p^{\frac{m-2}{2}}\right)/2$ \\   \hline
\end{tabular}
\end{center}}
where $\varepsilon_0=(-1)^{\frac{(p-1)m}{4}}$.
\end{table}
\end{description}
\end{theorem}
{\it Proof.} We only prove the case $p \equiv 1\pmod 4$, and the case $p \equiv 3\pmod 4$ can be shown by the similar way. There are two cases to be discussed.
\begin{description}
\item{\bf Case 1}: $v_2(m)=0$. From (\ref{111}), (\ref{112}) and Proposition \ref{pro:ww}, when $(a,b,c)$ runs over $(\mathbb{F}_{p^m}, \mathbb{F}_{p^m}, \mathbb{F}_p)$, the number for $\wt(\mathbf{c}(a,b,c))$ being $p^m$ is $p-1$.

We now determine the multiplicities of the other possible values of $\wt(\mathbf{c}(a,b,c))$ for $(a,b,c)$ running through $(\mathbb{F}_{p^m},\mathbb{F}_{p^m}, \mathbb{F}_p)$.
Define
\begin{equation*}\label{eq:quadraticformT}
N_{\epsilon_0,\epsilon_1}=\left|\left\{(a,b,c) \in (\mathbb{F}_{p^m},\mathbb{F}_{p^m},\mathbb{F}_p)\, |\, \wt(\mathbf{c}(a,b,c)) = (p-1)p^{m-1}+\epsilon_0p^{\frac{m-1}{2}}+\epsilon_1\right\}\right|,
\end{equation*}
where $\epsilon_0=\pm 1$, and $\epsilon_1\in\{0,1\}$. For any $a \in \mathbb{F}_{p^m}^*$, $a^{p^k} x^{p^{2k}} + ax $ is a permutation polynomial over $\mathbb{F}_{p^m}$ since $v_2(m)=0$.
When $b$ runs through $\mathbb{F}_{p^m}$, the corresponding solution $x_b$ of $a^{p^k} x^{p^{2k}} + ax + b^{p^k}=0$ runs through $\mathbb{F}_{p^m}$. So, the number of the pairs
$(c, x_b)$ such that $c={\rm Tr}_1^m\left(ax_b^{p^k+1}\right)$ is $p^m$ when $(b,c)$ runs over $(\mathbb{F}_{p^m}, \mathbb{F}_p)$.
Hence, from (\ref{111}), (\ref{112}) and Proposition \ref{pro:ww}, we have

\begin{equation*}
 \begin{split}
   N_{0,0}&=\left|\left\{(a,b,c) \in (\mathbb{F}_{p^m},\mathbb{F}_{p^m},\mathbb{F}_p)\,\,|\,\wt(\mathbf{c}(a,b,c))=(p-1)p^{m-1} \right\}\right|\\
&=\left|\left\{(a, b,c) \in (0, \mathbb{F}_{p^m},\mathbb{F}_p)\,\,|\,\wt(\mathbf{c}(a,b,c))=(p-1)p^{m-1} \right\}\right|\\
&+\left|\left\{(a,b,c) \in (\mathbb{F}_{p^m}^*,\mathbb{F}_{p^m},\mathbb{F}_p)\,\,|\,\wt(\mathbf{c}(a,b,c))=(p-1)p^{m-1} \right\}\right|\\
&=p(p^m-1)+\left|\left\{(a,b,c) \in (\mathbb{F}_{p^m}^*,\mathbb{F}_{p^m},\mathbb{F}_p)\,\,|\,{\rm Tr}_1^m(a)=0,\,\, c_0=0\right\}\right|\\
&=p(p^m-1)+p^m(p^{m-1}-1),
\end{split}
\end{equation*}
and
\begin{equation*}
\begin{split}
N_{0,1}&=\left|\left\{(a,b,c) \in (\mathbb{F}_{p^m},\mathbb{F}_{p^m},\mathbb{F}_p)\,\,|\,\wt(\mathbf{c}(a,b,c))=(p-1)p^{m-1}+1 \right\}\right|\\
&=\left|\left\{(a,b,c) \in (\mathbb{F}_{p^m},\mathbb{F}_{p^m},\mathbb{F}_p)\,\,|\,{\rm Tr}_1^m(a)\neq0,\,\, c_0=0\right\}\right|\\
&=p^m(p^m-p^{m-1}).
\end{split}
\end{equation*}

Recall that $c_0=c-{\rm Tr}_1^m\left(ax_b^{p^k+1}\right)$.
It is easy to see that $c_0$ runs through $\mathbb{F}_p$  if $c$ runs through $\mathbb{F}_p$ for any  $(a,b) \in (\mathbb{F}_{p^m}^*,\mathbb{F}_{p^m})$. So, when $(a,b,c)$ runs through $(\mathbb{F}_{p^m}^*,\mathbb{F}_{p^m}, \mathbb{F}_p)$, by Proposition \ref{pro:ww}, we have $N_{1,0}=N_{-1,0}$ and $N_{1,1}=N_{-1,1}$ since $\eta(ac_0)=\eta(a)\eta_0(c_0)$, where $\eta$ and $\eta_0$ are quadratic multiplicative characters of $\mathbb{F}_{p^m}^*$ and $\mathbb{F}_{p}^*$, respectively. Hence, from the first three Pless power moment identities, we have
\begin{equation*}
\begin{split}
\begin{cases}
N_{1,0}=N_{-1,0}=\frac{p^m(p^{m-1}-1)(p-1)}{2}, \\
N_{1,1}=N_{-1,1}=\frac{p^{2m-1}(p-1)^2}{2}.
\end{cases}
\end{split}
\end{equation*}

So, we obtain the weight distribution of $\mathcal{C}_k$ in Table \ref{Table1} for the case $v_2(m)=0$.

\item{\bf Case 2}: $0<v_2(m)\leq v_2(k)$. From (\ref{111}), (\ref{112}) and Proposition \ref{pro:ww}, when $a=0$ and $(b,c)$ runs over $(\mathbb{F}_{p^m}, \mathbb{F}_p)$, the number of $\wt(\mathbf{c}(a,b,c))$ being $p^m$ or $(p-1)p^{m-1}$ is $p-1$ or $p(p^m-1)$, respectively. We now determine the multiplicities of the other possible values of $\wt(\mathbf{c}(a,b,c))$ for $(a,b,c)$ running through $(\mathbb{F}_{p^m}^*,\mathbb{F}_{p^m}, \mathbb{F}_p)$.  Define
\begin{equation*}\label{eq:quadraticformT}
N_{\epsilon_0,\epsilon_1,\epsilon_2}=\left|\left\{(a,b,c) \in (\mathbb{F}_{p^m}^*,\mathbb{F}_{p^m},\mathbb{F}_p)\, |\, \wt(\mathbf{c}(a,b,c)) = (p-1)p^{m-1}+\epsilon_0p^{\frac{m}{2}}+\epsilon_1p^{\frac{m-2}{2}}+\epsilon_2\right\}\right|,
\end{equation*}
where $\epsilon_0\in\{0,1,-1\}$, $\epsilon_1\in \{1,-1\}$ and $\epsilon_2 \in \{0,1\}$. Then, by Lemma \ref{lemeven} we have
\begin{equation*}
\begin{split}
N_{1,-1,0}&=\left|\left\{(a,b,c) \in (\mathbb{F}_{p^m}^*,\mathbb{F}_{p^m},\mathbb{F}_p)\, |\, \wt(\mathbf{c}(a,b,c)) = (p-1)p^{m-1}+p^{\frac{m}{2}}-p^{\frac{m-2}{2}}\right\}\right|,\\
&=\left|\left\{(a,b,c) \in (\mathbb{F}_{p^m}^*,\mathbb{F}_{p^m},\mathbb{F}_p)\,\,|\,{\rm Tr}_1^m(a)=0,\,\,\eta(a)=1,\,\,c_0=0\right\}\right|\\
&=p^m\left(p^{m-1}-1-(p-1)p^{\frac{m-2}{2}}\right)/2,
\end{split}
\end{equation*}
and
 \begin{equation*}
\begin{split}
N_{1,-1,1}&=\left|\left\{(a,b,c) \in (\mathbb{F}_{p^m}^*,\mathbb{F}_{p^m},\mathbb{F}_p)\, |\, \wt(\mathbf{c}(a,b,c)) = (p-1)p^{m-1}+p^{\frac{m}{2}}-p^{\frac{m-2}{2}}+1\right\}\right|\\
&=\left|\left\{(a,b,c) \in (\mathbb{F}_{p^m}^*,\mathbb{F}_{p^m},\mathbb{F}_p)\,\,|\,{\rm Tr}_1^m(a)\neq0,\,\,\eta(a)=1,\,\,c_0=0\right\}\right|\\
&=p^m(p-1)\left(p^{m-1}+p^{\frac{m-2}{2}}\right)/2.
\end{split}
\end{equation*}
\end{description}
 By the similar calculations, we can get the values of $N_{-1,1,0}$, $N_{-1,1,1}$, $N_{0,-1,0}$, $N_{0,-1,1}$, $N_{0,1,0}$ and $N_{0,1,1}$. Then the  weight distribution of $\mathcal{C}_k$ is obtained in Table \ref{Table2} for
 the case $0<v_2(m)\leq v_2(k)$.

 \vskip 6pt
 From Table \ref{Table1}, Table \ref{Table2} and  the first five Pless power moments identities, we have that the dual of $\mathcal{C}_k$ has parameters $[p^m+1,p^m-2m,4]$ for the case $v_2(m)\leq v_2(k)$. By the Sphere Packing bound, we have
\begin{equation*}
\begin{split}
p^{p^m+1}\geq p^{p^m+1-(2m+1)}\left(\sum_{i=0}^{\lfloor\frac{{\rm d_H}(\mathcal{C}_k^{\perp})-1}{2}\rfloor}\left(
\begin{array}{cccc}
   p^m+1  \\
     i  \\
\end{array}
\right)(p-1)^i\right).
\end{split}
\end{equation*}
Hence, ${\rm d_H}(\mathcal{C}_k^{\perp})\leq 6$ if $p=3$ and ${\rm d_H}(\mathcal{C}_k^{\perp})\leq 4$ if $p\geq 5$. Therefore, when $p\geq 5$, $\mathcal{C}_k^{\perp}$ is optimal with respect to the Sphere Packing bound. $\square$

\begin{remark}
Let $k=0$ in Theorem \ref{theorem1}, then we obtain \cite[Theorem 16]{Hengar}. This means
that \cite[Theorem 16]{Hengar} can be seen as a special case of Theorem \ref{theorem1}.
\end{remark}
\begin{example}\label{example1}
Let $\mathcal{C}_k$ be the linear code in Theorem \ref{theorem1}.
\begin{description}
\item{(1)} Let $m=3$, $p=5$ and $k=1$. Then $\mathcal{C}_k$ has parameters $[126,7,95]$ and its dual has parameters $[126,119,4]$.
\item{(2)} Let $m=4$, $p=3$ and $k=0$.  Then $\mathcal{C}_k$ has parameters $[82,9,48]$ and its dual has parameters $[82,73,4]$.
\end{description}

All of these codes and the duals are optimal or almost optimal with respect to the tables of best codes known maintained at http://www.codetables.de.
\end{example}

\section{The weight distribution of $\mathcal{C}_k$ for $v_2(m)> v_2(k)$}

In this section, we always assume that  $v_2(m)> v_2(k)$, $d=\gcd(m,k)$, $\xi$ is a primitive element of $\mathbb{F}_{p^m}$ and $a_0=\xi^{\frac{p^m-1}{2(p^d-1)}}$, where $m$ and $k$ are positive integers. In the following, we determine the weight distribution of the code $\mathcal{C}_k$ for the case $v_2(m)> v_2(k)$, where $\mathcal{C}_k$ is defined in (\ref{code0}).  Recall that
\begin{equation*}
\begin{split}
S(a,b)=\sum_{x \in \mathbb{F}_{p^m}}\zeta_p^{{\rm Tr}_1^m\left(ax^{p^k+1}+bx\right)}.
\end{split}
\end{equation*}
The possible values of exponential sum $S(a,b)$ for the case $v_2(m)> v_2(k)$ have been given in \cite{Coulter1998} and \cite{Coulter2002}.

\begin{lemma}\cite{Coulter1998, Coulter2002}\label{lemeven1}
Let $a,b \in \mathbb{F}_{p^m}$ with $a\neq 0$. Then $S(a,b)=0$ unless the equation $a^{p^k}x^{p^{2k}}+ax+b^{p^k}=0$ is solvable. Let $\xi$ be a primitive element of $\mathbb{F}_{p^m}$ and $a_0=\xi^{\frac{p^m-1}{2(p^d-1)}}$. Assume $a^{p^k}x^{p^{2k}}+ax+b^{p^k}=0$ is solvable, then there are two possibilities.
\begin{description}
\item{(i)} If $a\neq a_0\xi^{s\left(p^d+1\right)}$ for any integer $s$, then the equation has a unique solution $x_b$ for any $b \in \mathbb{F}_{p^m}$, and
    $$S(a,b)=(-1)^{\frac{m}{2d}}p^{\frac{m}{2}}\zeta_p^{-{\rm Tr}_1^m\left(ax_b^{p^k+1}\right)}.$$
\item{(ii)} If $a= a_0\xi^{s\left(p^d+1\right)}$ for some integer $s$, then the equation is solvable if and only if ${\rm Tr}_{2d}^m\left(b\gamma^{-s}\right)=0$, where $\gamma \in \mathbb{F}_{p^m}^*$ is the unique element satisfying $\gamma^{\frac{p^k+1}{p^d+1}}=\xi$. In  such cases,
    $$S(a,b)=-(-1)^{\frac{m}{2d}}p^{\frac{m}{2}+d}\zeta_p^{-{\rm Tr}_1^m\left(ax_b^{p^k+1}\right)},$$
where $x_b$ is a solution of $a^{p^k}x^{p^{2k}}+ax+b^{p^k}=0$ and $d=\gcd(k,m)$.
\end{description}
\end{lemma}

Now, we determine the possible values of $\wt(\mathbf{c}(a,b,c))$ for $(a,b,c)$ running through $(\mathbb{F}_{p^m}^*,\mathbb{F}_{p^m}, \mathbb{F}_p)$.

\begin{proposition}\label{pro:ww1}
Let $c \in \mathbb{F}_p$ and the other notations be as in Lemma \ref{lemeven1}. Let $c_0=c-{\rm Tr}_1^m\left(ax_b^{p^k+1}\right)$.
If $a\neq a_0\xi^{\left(p^d+1\right)s}$ for any integer $s$, then the possible distinct values of $\wt(\mathbf{c}(a,b,c))$  are
\begin{equation*}
\begin{split}
\wt(\mathbf{c}(a,b,c))=\begin{cases}
(p-1)p^{m-1}+(-1)^{\frac{m}{2d}}p^{\frac{m-2}{2}}, & \text{if ${\rm Tr}_1^m(a)=0$ and $c_0\neq 0$,} \\
(p-1)p^{m-1}+(-1)^{\frac{m}{2d}}p^{\frac{m-2}{2}}+1, & \text{if ${\rm Tr}_1^m(a)\neq 0$ and $c_0\neq 0$,} \\
(p-1)p^{m-1}-(-1)^{\frac{m}{2d}}(p-1)p^{\frac{m-2}{2}}, & \text{if ${\rm Tr}_1^m(a)=0$ and $c_0=0$,} \\
(p-1)p^{m-1}-(-1)^{\frac{m}{2d}}(p-1)p^{\frac{m-2}{2}}+1, & \text{if ${\rm Tr}_1^m(a)\neq 0$ and $c_0=0$.} \\
\end{cases}
\end{split}
\end{equation*}
If $a=a_0\xi^{\left(p^d+1\right)s}$ for some integer $s$, then the possible distinct values of $\wt(\mathbf{c}(a,b,c))$ are
\begin{equation*}
\begin{split}
\wt(\mathbf{c}(a,b,c))=\begin{cases}
(p-1)p^{m-1}, & \text{if ${\rm Tr}_1^m(a)=0$ and ${\rm Tr}_{2d}^m(b\gamma^{-s})\neq0$}, \\
(p-1)p^{m-1}+1, & \text{if ${\rm Tr}_1^m(a)\neq 0$ and ${\rm Tr}_{2d}^m(b\gamma^{-s})\neq0$}, \\
(p-1)p^{m-1}-(-1)^{\frac{m}{2d}}p^{\frac{m+2d-2}{2}}, & \text{if ${\rm Tr}_1^m(a)=0$, ${\rm Tr}_{2d}^m(b\gamma^{-s})=0$ and $c_0\neq 0$,} \\
(p-1)p^{m-1}-(-1)^{\frac{m}{2d}}p^{\frac{m+2d-2}{2}}+1, & \text{if ${\rm Tr}_1^m(a)\neq 0$, ${\rm Tr}_{2d}^m(b\gamma^{-s})=0$ and $c_0\neq 0$,} \\
(p-1)p^{m-1}+(-1)^{\frac{m}{2d}}(p-1)p^{\frac{m+2d-2}{2}}, & \text{if ${\rm Tr}_1^m(a)=0$, ${\rm Tr}_{2d}^m(b\gamma^{-s})=0$ and $c_0=0$,} \\
(p-1)p^{m-1}+(-1)^{\frac{m}{2d}}(p-1)p^{\frac{m+2d-2}{2}}+1, & \text{if ${\rm Tr}_1^m(a)\neq 0$, ${\rm Tr}_{2d}^m(b\gamma^{-s})=0$ and $c_0=0$.} \\
\end{cases}
\end{split}
\end{equation*}
\end{proposition}

{\it Proof}. We only prove the case ${\rm Tr}_1^m(a)=0$. The case ${\rm Tr}_1^m(a)\neq 0$ can be proved similarly and omit the details here.

It is easy to see that $p^d+1\,\big|\, \frac{p^m-1}{p-1}$ since $v_2(m)> v_2(k)$ and $d=\gcd(m,k)$. Then, $z \in \mathbb{F}_p^*$ can be expressed as $z=\xi^{(p^d+1)i}$, where $i$ is a positive integer.  This means that $az= a_0\xi^{(p^d+1)s_1}$ if and only if $a= a_0\xi^{(p^d+1)s}$ for any $z \in \mathbb{F}_p^*$, where $s$, $s_1$ are positive integers. There are three cases.
\begin{description}
\item{\bf Case 1}: $a\neq a_0\xi^{(p^d+1)s}$ for any integer $s$. By Lemma \ref{lemeven1} we have
\begin{equation}\label{even1}
\begin{split}
\wt(\mathbf{c}(a,b,c))&=p^m-\frac{1}{p}\sum_{z \in \mathbb{F}_p}\sum_{x \in \mathbb{F}_{p^m}}\zeta_p^{z\left({\rm Tr}_1^m\left(ax^{p^k+1}+bx\right)+c\right)}\\
&=(p-1)p^{m-1}-\frac{1}{p}\sum_{z \in \mathbb{F}_p^*}\sum_{x \in \mathbb{F}_{p^m}}\zeta_p^{z\left({\rm Tr}_1^m\left(ax^{p^k+1}+bx\right)+c\right)}\\
&=(p-1)p^{m-1}-\frac{1}{p}\sum_{z \in \mathbb{F}_p^*}\zeta_p^{zc}\sum_{x \in \mathbb{F}_{p^m}}\zeta_p^{{\rm Tr}_1^m\left(zax^{p^k+1}+zbx\right)}\\
&=(p-1)p^{m-1}-(-1)^{\frac{m}{2d}}p^{\frac{m-2}{2}}\sum_{z \in \mathbb{F}_p^*}\zeta_p^{z\left(c-{\rm Tr}_1^m\left(ax_b^{p^k+1}\right)\right)}\\
&=\begin{cases}
(p-1)p^{m-1}+(-1)^{\frac{m}{2d}}p^{\frac{m-2}{2}}, & if\,\, c\neq{\rm Tr}_1^m\left(ax_b^{p^k+1}\right),\\
(p-1)p^{m-1}-(-1)^{\frac{m}{2d}}(p-1)p^{\frac{m-2}{2}}, & if\,\, c={\rm Tr}_1^m\left(ax_b^{p^k+1}\right).
\end{cases}
\end{split}
\end{equation}

\item{\bf Case 2}: $a=a_0\xi^{(p^d+1)s}$ for some integer $s$ and ${\rm Tr}_{2d}^m(b\gamma^{-s})\neq0$. From the third equality of (\ref{even1}) we have
\begin{equation*}
\wt(\mathbf{c}(a,b,c))=(p-1)p^{m-1}-\frac{1}{p}\sum_{z \in \mathbb{F}_p^*}\zeta_p^{zc}S(za,zb),
\end{equation*}
where
\begin{equation*}
\begin{split}
S(za,zb)=\sum_{x \in \mathbb{F}_{p^m}}\zeta_p^{{\rm Tr}_1^m\left(zax^{p^k+1}+zbx\right)}.
\end{split}
\end{equation*}
In this case, we have $S(za,zb)=S(a,b)=0$. Hence, $\wt(\mathbf{c}(a,b,c))=(p-1)p^{m-1}.$

\item{\bf Case 3}: $a=a_0\xi^{(p^d+1)s}$ for some integer $s$ and ${\rm Tr}_{2d}^m(b\gamma^{-s})=0$. By Lemma \ref{lemeven1} and the third equality~(\ref{even1}) we obtain
\begin{equation*}
\begin{split}
\wt(\mathbf{c}(a,b,c))
& =(p-1)p^{m-1}-\frac{1}{p}\sum_{z \in \mathbb{F}_p^*}\zeta_p^{zc}\sum_{x \in \mathbb{F}_{p^m}}\zeta_p^{{\rm Tr}_1^m\left(zax^{p^k+1}+zbx\right)}\\
&=(p-1)p^{m-1}+(-1)^{\frac{m}{2d}}p^{\frac{m+2d-2}{2}}\sum_{z \in \mathbb{F}_p^*}\zeta_p^{z\left(c-{\rm Tr}_1^m\left(ax_b^{p^k+1}\right)\right)}\\
&=\begin{cases}
(p-1)p^{m-1}-(-1)^{\frac{m}{2d}}p^{\frac{m+2d-2}{2}}, & \text{if $c\neq{\rm Tr}_1^m\left(ax_b^{p^k+1}\right)$,} \\
(p-1)p^{m-1}+(-1)^{\frac{m}{2d}}(p-1)p^{\frac{m+2d-2}{2}}, & \text{if $c={\rm Tr}_1^m\left(ax_b^{p^k+1}\right)$.} \\
\end{cases}
\end{split}
\end{equation*}
Combining with Case $1$, Case $2$ and Case $3$, the result follows. $\square$
\end{description}

In order to determine the weight distribution of $\mathcal{C}_k$ for the case $v_2(m)>v_2(k)$, we need the following lemma.
\begin{lemma}
Let $m$ and $d$ be two positive integers such that $d\, | \, m$ and $v_2(m)>v_2(d)$.  Let $\xi$ be a primitive element of $\mathbb{F}_{p^m}$, $a_0=\xi^{\frac{p^m-1}{2(p^d-1)}}$ and $a=a_0\xi^{(p^d+1)s}$ for some integer $s$. Let $N(a)$ be the number of $a \in \mathbb{F}_{p^m}$ such that ${\rm Tr}_1^m(a)=0$, then
\begin{equation*}
\begin{split}
&N(a)=\begin{cases}
\frac{p^{m-1}+p^{\frac{m+2d}{2}}-p^{\frac{m+2d-2}{2}}-1}{p^d+1}, & \text{if $v_2(m)=v_2(d)+1$,}\\
\frac{p^{m-1}-p^{\frac{m+2d}{2}}+p^{\frac{m+2d-2}{2}}-1}{p^d+1}, & \text{if $v_2(m)>v_2(d)+1$.}
\end{cases}
\end{split}
\end{equation*}
\end{lemma}
{\it Proof}. Let $G$ be the cyclic group generated by $\xi^{p^d+1}$. Obviously, $p^d+1 \, | \, p^m-1$ since $d\, | \, m$ and $v_2(m)>v_2(d)$, then by Lemma \ref{lem:quadraticsum} we have

\begin{equation*}
\begin{split}
N(a)&= \frac{1}{p}\sum_{z \in \mathbb{F}_p}\sum_{s=0}^{\frac{p^m-1}{p^d+1}-1}\zeta_p^{z{\rm Tr}_1^m\left(a_0\xi^{\left(p^d+1\right)s}\right)} \\
&=\frac{1}{p}\sum_{z \in \mathbb{F}_p}\left(\frac{1}{p^d+1}\sum_{x \in \mathbb{F}_{p^m}^*}\zeta_p^{z{\rm Tr}_1^m\left(a_0x^{p^d+1}\right)}\right)\\
&=\frac{p^m-p}{p(p^d+1)}+\frac{1}{p(p^d+1)}\sum_{z \in \mathbb{F}_p^*}\sum_{x \in \mathbb{F}_{p^m}}\zeta_p^{z{\rm Tr}_1^m\left(a_0x^{p^d+1}\right)}\\
&=\frac{p^m-p}{p(p^d+1)}+\frac{1}{p(p^d+1)}\sum_{z \in \mathbb{F}_p^*}\eta^r(z)\sum_{x \in \mathbb{F}_{p^m}}\zeta_p^{{\rm Tr}_1^m\left(a_0x^{p^d+1}\right)}\\
&=\frac{p^m-p}{p(p^d+1)}+\frac{p-1}{p(p^d+1)}\sum_{x \in \mathbb{F}_{p^m}}\zeta_p^{{\rm Tr}_1^m\left(a_0x^{p^d+1}\right)},\\
\end{split}
\end{equation*}
where $r$ is the rank of the quadratic form ${\rm Tr}_1^m\left(a_0x^{p^d+1}\right)$.  In the last equality, we used the fact that the rank of the quadratic form ${\rm Tr}_1^m\left(a_0x^{p^d+1}\right)$ is even, which has been given in \cite{Coulter19980,Draper2007}. Clearly,
\begin{equation*}
\begin{split}
a_0^{\frac{(p^d-1)(p^m-1)}{p^{\gcd(2d,m)}-1}}=\xi^{{\frac{p^m-1}{2(p^d-1)}}\cdot{\frac{(p^d-1)(p^m-1)}{p^{\gcd(2d,m)}-1}}}=\begin{cases}
-1, & \text{if $v_2(m)=v_2(d)+1$,}\\
1, & \text{if $v_2(m)>v_2(d)+1$.}
\end{cases}
\end{split}
\end{equation*}
Hence, by Lemma \ref{cor:quadraticsum} we have
 \begin{equation*}
\begin{split}
N(a)&=\begin{cases}
\frac{p^{m-1}+p^{\frac{m+2d}{2}}-p^{\frac{m+2d-2}{2}}-1}{p^d+1}, & \text{if $v_2(m)=v_2(d)+1$,}\\
\frac{p^{m-1}-p^{\frac{m+2d}{2}}+p^{\frac{m+2d-2}{2}}-1}{p^d+1}, & \text{if $v_2(m)>v_2(d)+1$.}
\end{cases}
\end{split}
\end{equation*}
 The result follows. $\square$

\vskip 6pt
With above preparations, we now give the weight distribution of $\mathcal{C}_k$ for the case $v_2(m)>v_2(k)$.
\begin{theorem}\label{thm2}
 Let $m, k$ be positive integers with $v_2(m)>v_2(k)$, $d=\gcd(k,m)$ and $\mathcal{C}_k$ be the linear code defined in (\ref{code0}). The following statements hold.
 \begin{description}
\item{(1)}  If $v_2(m)=v_2(k)+1$ $(m\neq 2k)$, then $\mathcal{C}_k$ is a $\left[p^m+1, 2m+1,(p-1)\left(p^{m-1}-p^{\frac{m+2d-2}{2}}\right)\right]$ code with  the weight distribution in Table \ref{Table3}. Its dual has parameters $[p^m+1,p^m-2m,4]$, which is optimal with respect to the Sphere Packing bound for $p>3$.
\begin{table}[h]
{\caption{\rm {  The weight distribution of $\mathcal{C}_k$ for $v_2(m)>v_2(k)$ with $m\neq 2k$ } }\label{Table3}
\begin{center}
\begin{tabular}{cccc}\hline
     Weight & Multiplicity \\\hline
  $0$ & $1$  \\
 $p^m$ & $p-1$ \\
 $(p-1)p^{m-1}$ & $\left(p^{m-d}-p^{m-2d}\right)\left(p^{m}+p^{\frac{m+2d+2}{2}}-p^{\frac{m+2d}{2}}-p\right)+p\left(p^m-1\right)$ \\
 $(p-1)p^{m-1}+1$ & $(p-1)\left(p^{m-d}-p^{m-2d}\right)\left(p^m-p^{\frac{m+2d}{2}}\right)$ \\
 $(p-1)p^{m-1}+(-1)^{\frac{m}{2d}}p^{\frac{m-2}{2}}$ & $(p-1)(p^{2m+d-1}-p^{\frac{3m+2d}{2}}+p^{\frac{3m+2d-2}{2}}-p^{m+d})/(p^d+1)$ \\
 $(p-1)p^{m-1}+(-1)^{\frac{m}{2d}}p^{\frac{m-2}{2}}+1$ & $(p-1)^2\left(p^{2m+d-1}+p^{\frac{3m+2d-2}{2}}\right)/(p^d+1)$ \\
 $(p-1)\left(p^{m-1}-(-1)^{\frac{m}{2d}}p^{\frac{m-2}{2}}\right)$ & $\left(p^{2m+d-1}-p^{\frac{3m+2d}{2}}+p^{\frac{3m+2d-2}{2}}-p^{m+d}\right)/(p^d+1)$ \\
    $(p-1)\left(p^{m-1}-(-1)^{\frac{m}{2d}}p^{\frac{m-2}{2}}\right)+1$ & $(p-1)\left(p^{2m+d-1}+p^{\frac{3m+2d-2}{2}}\right)/(p^d+1)$ \\
     $(p-1)p^{m-1}-(-1)^{\frac{m}{2d}}p^{\frac{m+2d-2}{2}}$ & $\left(p^{m-1}+p^{\frac{m+2d}{2}}-p^{\frac{m+2d-2}{2}}-1\right)(p-1)p^{m-2d}/(p^d+1)$ \\
  $(p-1)p^{m-1}-(-1)^{\frac{m}{2d}}p^{\frac{m+2d-2}{2}}+1$ & $(p-1)^2\left(p^{2m-2d-1}-p^{\frac{3m-2d-2}{2}}\right)/(p^d+1)$ \\
     $(p-1)\left(p^{m-1}+(-1)^{\frac{m}{2d}}p^{\frac{m+2d-2}{2}}\right)$ & $\left(p^{m-1}+p^{\frac{m}{2}+d}-p^{\frac{m}{2}+d-1}-1\right)p^{m-2d}/(p^d+1)$ \\
  $(p-1)\left(p^{m-1}+(-1)^{\frac{m}{2d}}p^{\frac{m+2d-2}{2}}\right)+1$ & $(p-1)\left(p^{2m-2d-1}-p^{\frac{3m-2d-2}{2}}\right)/(p^d+1)$ \\   \hline
\end{tabular}
\end{center}}
\end{table}
\item{(2)}  If $v_2(m)>v_2(k)+1$, then $\mathcal{C}_k$ is a $\left[p^m+1, 2m+1,(p-1)p^{m-1}-p^{\frac{m+2d-2}{2}}\right]$ code with the weight distribution in Table \ref{Table3}. Its dual has parameters $[p^m+1,p^m-2m,4]$, which is optimal with respect to the Sphere Packing bound for $p>3$.

\item{(3)}  If $m=2k$, then $\mathcal{C}_k$ is a $\left[p^m+1, \frac{3m}{2}+1,(p-1)p^{m-1}-p^{\frac{m-2}{2}}\right]$ code with the weight distribution in Table \ref{Table4}. Its dual has parameters
$[p^m+1,p^m-\frac{3m}{2},4]$, which is optimal with respect to the Sphere Packing bound for $p>3$.
\begin{table}[h]
{\caption{\rm   The weight distribution of $\mathcal{C}_k$ for $m=2k$   }\label{Table4}
\begin{center}
\begin{tabular}{cccc}\hline
     Weight & Multiplicity \\\hline
  $0$ & $1$  \\
 $p^m$ & $p-1$ \\
 $(p-1)p^{m-1}$ & $p^{m+1}-p$ \\
 $(p-1)p^{m-1}-p^{\frac{m-2}{2}}$ & $(p-1)(p^{\frac{3m}{2}-1}-p^{m})$ \\
 $(p-1)p^{m-1}-p^{\frac{m-2}{2}}+1$ & $(p-1)^2p^{\frac{3m}{2}-1}$ \\
 $(p-1)\left(p^{m-1}+p^{\frac{m-2}{2}}\right)$ & $p^{\frac{3m}{2}-1}-p^{m}$ \\
    $(p-1)\left(p^{m-1}+p^{\frac{m-2}{2}}\right)+1$ & $(p-1)p^{\frac{3m}{2}-1}$ \\   \hline
\end{tabular}
\end{center}}
\end{table}
\end{description}
\end{theorem}
{\it Proof.} We only prove the case $v_2(m)=v_2(k)+1$. The case $v_2(m)>v_2(k)+1$ can be shown by the similar way.
\vskip 6pt
The multiplicities of all possible values of $\wt(\mathbf{c}(a,b,c))$ for $(a,b,c)$ running through $(0,\mathbb{F}_{p^m}, \mathbb{F}_p)$ have been given in (\ref{111}), (\ref{112}). In the following, we only consider the multiplicities of all possible values of $\wt(\mathbf{c}(a,b,c))$ for $(a,b,c)$ running through $(\mathbb{F}_{p^m}^*,\mathbb{F}_{p^m}, \mathbb{F}_p)$.
There are two cases to be discussed.
\begin{description}
\item{\bf Case 1}: $a\neq a_0\xi^{(p^d+1)s}$ for any integer $s$. Define
 \begin{equation*}
\begin{split}
H_{\epsilon_0,\epsilon_1}=&\big|\big\{ (a, b,c) \in (\bF_{p^m}^*, \mathbb{F}_{p^m},\mathbb{F}_p)\,\,|\,\, a\neq a_0\xi^{(p^d+1)s} \,\, \text{for any $s$}, \\
&\,\,\,\,\,\,\,\,\,\, \wt(\mathbf{c}(a,b,c)) = (p-1)p^{m-1}+\epsilon_0p^{\frac{m}{2}}-p^{\frac{m-2}{2}}+\epsilon_1\big\}\big|,
\end{split}
\end{equation*}
where $\epsilon_0,\epsilon_1\in\{0,1\}$. By Proposition \ref{pro:ww1} we have
\begin{equation*}
\begin{split}
H_{0,0}&=\left|\left\{(a, b,c) \in (\bF_{p^m}^*, \mathbb{F}_{p^m},\mathbb{F}_p)\, |\, a\neq a_0\xi^{(p^d+1)s} \,\, \text{for any $s$}, \wt(\mathbf{c}(a,b,c)) = (p-1)p^{m-1}-p^{\frac{m-2}{2}}\right\}\right|\\
&=\left|\left\{(a,b,c) \in (\mathbb{F}_{p^m}^*,\mathbb{F}_{p^m},\mathbb{F}_p)\, |\, {\rm Tr}_1^m(a)=0,\,\,  c_0\neq 0 \right\}\right|\\
&-\left|\left\{(a, b,c) \in (\bF_{p^m}^*, \mathbb{F}_{p^m},\mathbb{F}_p)\, |\, a=a_0\xi^{(p^d+1)s} \,\,\text{for some $s$}, {\rm Tr}_1^m(a)=0,\,\,  c_0\neq 0 \right\}\right|\\
&=(p^{m-1}-1)(p-1)p^m-\frac{\left(p^{m-1}+p^{\frac{m+2d}{2}}-p^{\frac{m+2d-2}{2}}-1\right)(p-1)p^m}{p^d+1}\\
&=\frac{(p-1)\left(p^{2m+d-1}-p^{\frac{3m+2d}{2}}+p^{\frac{3m+2d-2}{2}}-p^{m+d}\right)}{p^d+1},
\end{split}
\end{equation*}
and
\begin{equation*}
\begin{split}
H_{0,1}&=\left|\left\{(a, b,c) \in (\bF_{p^m}^*, \mathbb{F}_{p^m},\mathbb{F}_p)\,\, |\, a\neq a_0\xi^{(p^d+1)s} \,\, \text{for any $s$}, \wt(\mathbf{c}(a,b,c)) = (p-1)p^{m-1}-p^{\frac{m-2}{2}}+1\right\}\right|\\
&=\left|\left\{(a,b,c) \in (\mathbb{F}_{p^m}^*,\mathbb{F}_{p^m},\mathbb{F}_p)\, |\, {\rm Tr}_1^m(a)\neq0,\,\,  c_0\neq 0 \right\}\right|\\
&-\left|\left\{ (a, b,c) \in (\bF_{p^m}^*\mathbb{F}_{p^m},\mathbb{F}_p)\, |\,a=a_0\xi^{(p^d+1)s} \,\,\text{for some $s$},\,\, {\rm Tr}_1^m(a)\neq 0,\,\,  c_0\neq 0 \right\}\right|\\
&=\left|\left\{(a,b,c) \in (\mathbb{F}_{p^m}^*,\mathbb{F}_{p^m},\mathbb{F}_p)\, |\, {\rm Tr}_1^m(a)\neq0,\,\,  c_0\neq 0 \right\}\right|\\
&-\left|\left\{ (a, b,c) \in (\bF_{p^m}^*, \mathbb{F}_{p^m},\mathbb{F}_p)\, |\,  a=a_0\xi^{(p^d+1)s} \,\,\text{for some $s$},\,\, c_0\neq 0 \right\}\right|\\
&+\left|\left\{  (a, b,c) \in (\mathbb{F}_{p^m},\mathbb{F}_p)\, |\,a=a_0\xi^{(p^d+1)s} \,\,\text{for some $s$},\,\, {\rm Tr}_1^m(a)=0,\,\, c_0\neq 0 \right\}\right|\\
&=(p-1)^2p^{2m-1}-\frac{(p^m-1)(p-1)p^m}{p^d+1}+\frac{\left(p^{m-1}+p^{\frac{m+2d}{2}}-p^{\frac{m+2d-2}{2}}-1\right)(p-1)p^m}{p^d+1}\\
&= \frac{(p-1)^2\left(p^{2m+d-1}+p^{\frac{3m+2d-2}{2}}\right)}{p^d+1}.
\end{split}
\end{equation*}

 By the similar calculations, we can get the values of $H_{1,0}$ and $H_{1,1}$.

\item{\bf Case 2}: $a=a_0\xi^{(p^d+1)s}$ for some integer $s$.
Define
 \begin{equation*}
\begin{split}
T_{\epsilon_0,\epsilon_1,\epsilon_2}=&\big|\big\{ (a, b,c) \in (\bF_{p^m}^*, \mathbb{F}_{p^m},\mathbb{F}_p)\, |\, a=a_0\xi^{(p^d+1)s} \text{ for some $s$}, \\
&\,\,\,\,\,\,\,\,\,\, \wt(\mathbf{c}(a,b,c)) = (p-1)p^{m-1}-\epsilon_0p^{\frac{m+2d}{2}}+\epsilon_1p^{\frac{m+2d-2}{2}}+\epsilon_2\big\}\big|,
\end{split}
\end{equation*}
where $\epsilon_0,\epsilon_1,\epsilon_2\in\{0,1\}$. By Proposition \ref{pro:ww1} we have

\begin{equation*}
\begin{split}
T_{0,0,0}&=\left|\left\{(a, b,c) \in (\bF_{p^m}^*, \mathbb{F}_{p^m},\mathbb{F}_p)\, |\, a=a_0\xi^{(p^d+1)s} \text{ for some $s$},  \wt(\mathbf{c}(a,b,c)) = (p-1)p^{m-1}\right\}\right|\\
&=p\left|\left\{ (a, b) \in ( \bF_{p^m}^*, \mathbb{F}_{p^m}) \, |\, a=a_0\xi^{(p^d+1)s} \,\,\text{for some $s$},\, {\rm Tr}_1^m(a)=0,\, {\rm Tr}_{2d}^m(b\gamma^{-s})\neq0 \right\}\right|\\
&=\frac{\left(p^m-p^{m-2d}\right)\left(p^{m}+p^{\frac{m+2d+2}{2}}-p^{\frac{m+2d}{2}}-p\right)}{p^d+1}\\
&=\left(p^{m-d}-p^{m-2d}\right)\left(p^{m}+p^{\frac{m+2d+2}{2}}-p^{\frac{m+2d}{2}}-p\right),\\
\end{split}
\end{equation*}

\begin{equation*}
\begin{split}
T_{0,0,1}&=\left|\left\{ (a, b,c) \in (\bF_{p^m}^*, \mathbb{F}_{p^m},\mathbb{F}_p)\, |\,a=a_0\xi^{(p^d+1)s} \text{ for some $s$}, \wt(\mathbf{c}(a,b,c)) = (p-1)p^{m-1}+1\right\}\right|\\
&=p\left|\left\{(a, b)\in (\bF_{p^m}^*, \mathbb{F}_{p^m})\, |\,  a=a_0\xi^{(p^d+1)s} \,\,\text{for some $s$},\, {\rm Tr}_1^m(a)\neq 0,\,  {\rm Tr}_{2d}^m(b\gamma^{-s})\neq0 \right\}\right|\\
&=p\left|\left\{ (a, b) \in ( \bF_{p^m}^*, \mathbb{F}_{p^m})\, |\, a=a_0\xi^{(p^d+1)s} \,\,\text{for some $s$},\, {\rm Tr}_{2d}^m(b\gamma^{-s})\neq0 \right\}\right|-T_{0,0,0}\\
&=\frac{\left(p^{m}-p^{m-2d}\right)\left(p^{m+1}-p\right)}{p^d+1}-\frac{\left(p^m-p^{m-2d}\right)\left(p^{m}+p^{\frac{m+2d+2}{2}}-p^{\frac{m+2d}{2}}-p\right)}{p^d+1}\\
&=(p-1)\left(p^{m-d}-p^{m-2d}\right)\left(p^m-p^{\frac{m}{2}+d}\right),
\end{split}
\end{equation*}
and
\begin{equation*}
\begin{split}
T_{0,1,1}&=\left|\left\{ (a, b,c) \in (\bF_{p^m}^*, \mathbb{F}_{p^m},\mathbb{F}_p)\, |\,a=a_0\xi^{(p^d+1)s} \,\text{for some $s$}, \, \wt(\mathbf{c}(a,b,c)) = (p-1)p^{m-1}+p^{\frac{m+2d-2}{2}}+1\right\}\right|\\
&=\left|\left\{ (a, b,c) \in (\bF_{p^m}^*, \mathbb{F}_{p^m},\mathbb{F}_p)\, |\, a=a_0\xi^{(p^d+1)s} \,\text{for some $s$},\, {\rm Tr}_1^m(a)\neq0,\,\,{\rm Tr}_{2d}^m(b\gamma^{-s})=0,\,\,  c_0\neq 0 \right\}\right|\\
&=\left|\left\{(a, b,c) \in (\bF_{p^m}^*, \mathbb{F}_{p^m},\mathbb{F}_p)\, |\,a=a_0\xi^{(p^d+1)s} \text{ for some $s$},\, {\rm Tr}_{2d}^m(b\gamma^{-s})=0,\,\,   c_0\neq 0 \right\}\right|\\
&-\left|\left\{ (a,b,c) \in (\mathbb{F}_{p^m},\mathbb{F}_p)\, |\,a=a_0\xi^{(p^d+1)s} \,\,\text{for some $s$},\, {\rm Tr}_1^m(a)=0,\,\,{\rm Tr}_{2d}^m(b\gamma^{-s})=0,\,\,  c_0\neq 0 \right\}\right|\\
&=\frac{(p^{m}-1)(p-1)p^{m-2d}}{p^d+1}-\frac{\left(p^{m-1}+p^{\frac{m+2d}{2}}-p^{\frac{m+2d-2}{2}}-1\right)(p-1)p^{m-2d}}{p^d+1}\\
&=\frac{(p-1)^2\left(p^{2m-2d-1}-p^{\frac{3m-2d-2}{2}}\right)}{p^d+1}.
\end{split}
\end{equation*}

\end{description}
By the similar calculations, we get the values of $T_{1,1,1}$, $T_{1,1,0}$ and $T_{0,1,0}$. Combining with (\ref{111}), (\ref{112}), Case $1$ and Case $2$, we  get Table \ref{Table3}.
\vskip 6pt

When $v_2(m)\geq v_2(k)$, from the first five Pless power moment identities and the weight distribution of $\mathcal{C}_k$, we obtain that the dual of $\mathcal{C}_k$ is a $[p^m+1,p^m-2m,4]$ code,
which is optimal with respect to the Sphere Packing bound for the case $p\geq 5$.

\vskip 6pt
When $m=2k$, i.e., $m=2d$, from Table \ref{Table3}, we know that the Hamming weight $0$ occurs $p^{\frac{m}{2}}$ times if $(a,b,c)$ runs through $(\mathbb{F}_{p^m},\mathbb{F}_{p^m}, \mathbb{F}_p)$. This means that every codeword in $\mathcal{C}_k$ repeats $p^{\frac{m}{2}}$ times. Hence, in this case, $\mathcal{C}_k$ is degenerate and its dimension is $\frac{3m}{2}$.  Substituting $d=\frac{m}{2}$ to Table \ref{Table3} and dividing each frequency by $p^\frac{m}{2}$, we get Table \ref{Table4}. From the first five Pless power moments identities and the weight distribution of $\mathcal{C}_k$,  we have the dual of $\mathcal{C}_k$ is a $[p^m+1,p^m-\frac{3m}{2},4]$ code, which is optimal with respect to the Sphere Packing bound for the case $p\geq 5$. $\square$

\begin{example}\label{example2}
Let $\mathcal{C}_k$ be the linear code in Theorem \ref{thm2}.
\begin{description}
\item{(1)} Let $m=2$, $p=5$ and $k=1$. Then $\mathcal{C}_k$ has parameters $[26,4,19]$ and its dual has parameters $[26,22,4]$.
\item{(2)} Let $m=4$, $p=3$ and $k=1$.  Then $\mathcal{C}_k$ has parameters $[82,9,45]$ and its dual has parameters $[82,73,4]$.
\end{description}

The duals of these codes are optimal or almost optimal with respect to the tables of best codes known maintained at http://www.codetables.de.
\end{example}

\section{The punctured code of $\mathcal{C}_k$}

In this section, we study the linear code
\begin{equation}\label{eqbarC}
\begin{split}
\mathcal{\bar{C}}_k=\left\{\left( {\rm Tr}_1^m\left(ax^{p^k+1}+bx\right)+c\right)_{x \in \mathbb{F}_{p^m}}: \, a,b \in \mathbb{F}_{p^m}, c \in \mathbb{F}_p\right\}.
\end{split}
\end{equation}
 Note that $\mathcal{\bar{C}}_k$ can be viewed as a punctured code of  $\mathcal{C}_k$ in Section $3$ and Section $4$. Hence, the weight distribution of $\mathcal{\bar{C}}_k$ can be easily derived from that of $\mathcal{C}_k$. We here give the weight distribution of $\mathcal{\bar{C}}_k$ and the parameters of the dual of  $\mathcal{\bar{C}}_k$.

\begin{theorem}\label{thm3}
 Let $d=\gcd(k,m)$ and $\mathcal{\bar{C}}_k$ be the linear code defined in (\ref{eqbarC}). The following statements hold.
 \begin{description}
\item{(1)}   If $v_2(m)=0$, then $\mathcal{\bar{C}}_k$ is a $\left[p^m, 2m+1, (p-1)p^{m-1}- p^{\frac{m-1}{2}}\right]$ code with the weight distribution in Table \ref{Table5}. Its dual has parameters $[p^m,p^m-2m-1,4]$ if $p>3$, and its dual has parameters $[p^m,p^m-2m-1,5]$ if $p=3$ and $m>1$.
\begin{table}[h]
{\caption{\rm   The weight distribution of $\mathcal{\bar{C}}_k$ for $v_2(m)=0$   }\label{Table5}
\begin{center}
\begin{tabular}{cccc}\hline
     Weight & Multiplicity \\\hline
  $0$ & $1$  \\
 $p^m$ & $p-1$ \\
 $(p-1)p^{m-1}$ & $(p+p^m)(p^m-1)$ \\
 $(p-1)p^{m-1}\pm  p^{\frac{m-1}{2}}$ & $p^m(p^m-1)(p-1)/2$ \\   \hline
\end{tabular}
\end{center}}

\end{table}
\item{(2)}  If $0<v_2(m)\leq v_2(k)$, then $\mathcal{\bar{C}}_k$ is a $\left[p^m, 2m+1,(p-1)\left(p^{m-1}-p^{\frac{m-2}{2}}\right)\right]$ code with the weight distribution in Table \ref{Table6}. Its dual has parameters $[p^m,p^m-2m-1,4]$ if $p>3$, and its dual has parameters $[p^m,p^m-2m-1,5]$ if $p=3$.
\begin{table}[h]
{\caption{\rm   The weight distribution of $\mathcal{\bar{C}}_k$ for $0<v_2(m)\leq v_2(k)$   }\label{Table6}
\begin{center}
\begin{tabular}{cccc}\hline
     Weight & Multiplicity \\\hline
  $0$ & $1$  \\
 $p^m$ & $p-1$ \\
  $(p-1)p^{m-1}$ & $p(p^m-1)$ \\
  $(p-1)\left(p^{m-1}\pm p^{\frac{m-2}{2}}\right)$ & $(p^{2m}-p^m)/2$ \\
  $(p-1)p^{m-1}\pm p^{\frac{m-2}{2}}$ & $(p-1)(p^{2m}-p^m)/2$ \\   \hline
\end{tabular}
\end{center}}
\end{table}
\item{(3)}  If $v_2(m)=v_2(k)+1$ $(m\neq 2k)$, then $\mathcal{\bar{C}}_k$ is a $\left[p^m, 2m+1,(p-1)\left(p^{m-1}-p^{\frac{m+2d-2}{2}}\right)\right]$ code with the weight distribution in Table \ref{Table7}. Its dual has parameters $[p^m,p^m-2m-1,4]$.

\item{(4)}  If $v_2(m)> v_2(k)+1$, then $\mathcal{\bar{C}}_k$ is a $\left[p^m, 2m+1,(p-1)p^{m-1}-p^{\frac{m+2d-2}{2}}\right]$ code with the weight distribution in Table \ref{Table7}. Its dual has parameters $[p^m,p^m-2m-1,4]$.
\begin{table}[h]
{\caption{\rm   The weight distribution of $\mathcal{\bar{C}}_k$ for $v_2(m)>v_2(k)$   }\label{Table7}
\begin{center}
\begin{tabular}{cccc}\hline
     Weight & Multiplicity \\\hline
  $0$ & $1$  \\
 $p^m$ & $p-1$ \\
 $(p-1)p^{m-1}$ & $p^{2m-d+1}-p^{2m-2d+1}+p^{m+1}-p^{m-d+1}+p^{m-2d+1}-p$ \\
 $(p-1)p^{m-1}+(-1)^{\frac{m}{2d}}p^{\frac{m-2}{2}}$ & $(p-1)(p^{2m+d}-p^{m+d})/(p^d+1)$ \\
 $(p-1)\left(p^{m-1}-(-1)^{\frac{m}{2d}}p^{\frac{m-2}{2}}\right)$ & $(p^{2m+d}-p^{m+d})/(p^d+1)$ \\
 $(p-1)p^{m-1}-(-1)^{\frac{m}{2d}}p^{\frac{m+2d-2}{2}}$ & $(p-1)(p^{2m-2d}-p^{m-2d})/(p^d+1)$ \\
 $(p-1)\left(p^{m-1}+(-1)^{\frac{m}{2d}}p^{\frac{m+2d-2}{2}}\right)$ & $(p^{2m-2d}-p^{m-2d})/(p^d+1)$ \\   \hline
\end{tabular}
\end{center}}
\end{table}
\item{(5)}  If $m=2k$, then $\mathcal{\bar{C}}_k$ is a $\left[p^m, \frac{3m}{2}+1,(p-1)p^{m-1}-p^{\frac{m-2}{2}}\right]$ code with the weight distribution in Table \ref{Table8}. Its dual has parameters $[p^m,p^m-\frac{3m}{2}-1,4]$.
    \begin{table}[h]
{\caption{\rm   The weight distribution of $\mathcal{\bar{C}}_k$ for $m=2k$   }\label{Table8}
\begin{center}
\begin{tabular}{cccc}\hline
     Weight & Multiplicity \\\hline
  $0$ & $1$  \\
 $p^m$ & $p-1$ \\
 $(p-1)p^{m-1}$ & $p^{m+1}-p$ \\
 $(p-1)p^{m-1}-p^{\frac{m-2}{2}}$ & $(p-1)(p^{\frac{3m}{2}}-p^m)$ \\
 $(p-1)\left(p^{m-1}+p^{\frac{m-2}{2}}\right)$ & $p^{\frac{3m}{2}}-p^m$ \\   \hline
\end{tabular}
\end{center}}
\end{table}
\item{(6)}  The dual of $\mathcal{\bar{C}}_k$ is an optimal code with respect to the Sphere Packing bound if $p>3$, and the dual of $\mathcal{\bar{C}}_k$ is an optimal ternary code for the case $v_2(m)\leq v_2(k)$ if $p=3$ and $m>1$.
\end{description}
\end{theorem}
{\it Proof.} The weight distribution of $\mathcal{\bar{C}}_k$ can be derived directly from Theorem \ref{theorem1} and Theorem \ref{thm2}. In the following, we will show the minimum Hamming distance of the dual of $\mathcal{\bar{C}}_k$. There are two cases.
\begin{description}
\item{{\bf Case 1}: $v_2(m)> v_2(k)$}. By the first five Pless power moment identities,  we get that ${\rm d_H}(\mathcal{\bar{C}}^{\perp}_k)=4$, which is optimal with respect to the Sphere Packing bound for the case $p>3$.
\item{{\bf Case 2}: $v_2(m)\leq v_2(k)$}. By the first five Pless power moment identities,  we obtain that ${\rm d_H}(\mathcal{\bar{C}}^{\perp}_k)=4$ if $p> 3$ and ${\rm d_H}(\mathcal{\bar{C}}^{\perp}_k)>4$ if $p=3$. Clearly, the linear code $\mathcal{\bar{C}}^{\perp}_k$ is optimal with respect to the Sphere Packing bound if $p> 3$. We now show ${\rm d_H}(\mathcal{\bar{C}}^{\perp}_k)=5$ if $p=3$ and $m>1$ (If $m=1$ and $p=3$, then $\mathcal{\bar{C}}_k$ is a $[3,3,1]$ code and $\mathcal{\bar{C}}^{\perp}_k=\mathbf{0}$).

    On on hand, if $p=3$, we have ${\rm d_H}(\mathcal{\bar{C}}^{\perp}_k)\leq 6$ by the Sphere Packing bound. On the other hand, assume that there exists a ternary linear code with parameters $[3^m, 3^m-2m-1,6]$. Applying Lemma \ref{bound2}, we have $q=3$, $n=3^m$, $t=3^m-5$, $r=2$, and
    $$3^{3^m-2m-1}\leq \frac{3^{3^m-1}}{1+2(3^m-1)^2},$$
    which is impossible if $m>1$. Hence, ${\rm d_H}(\mathcal{\bar{C}}^{\perp}_k)=5$ if $p=3$ and $m>1$, which is optimal with respect to Lemma \ref{bound2}. $\square$
\end{description}

\begin{remark}
 It is clear that $\mathcal{\bar{C}}_k$ has fewer distinct weights than $\mathcal{C}_k$ and the dual of $\mathcal{\bar{C}}_k$ is optimal for many cases. This means that it is worthwhile to study the linear code $\mathcal{\bar{C}}_k$.
\end{remark}

\begin{example}\label{example3}
Let $\mathcal{\bar{C}}_k$ be the linear code in Theorem \ref{thm3}.
\begin{description}
\item{(1)} Let $m=3$, $p=5$ and $k=1$. Then $\mathcal{C}_k$ has parameters $[125,7,95]$ and its dual has parameters $[125,118,4]$.
\item{(2)} Let $m=3$, $p=3$ and $k=1$.  Then $\mathcal{C}_k$ has parameters $[27,7,15]$ and its dual has parameters $[27,20,5]$.
\item{(3)} Let $m=4$, $p=3$ and $k=2$.  Then $\mathcal{C}_k$ has parameters $[81,7,51]$ and its dual has parameters $[81,74,4]$.
\end{description}

All of these codes and the duals are optimal with respect to the tables of best codes known maintained at http://www.codetables.de.
\end{example}

\section{Conclusion }

This paper continued the work of Heng and Ding in~\cite{Hengar}, and investigated more subfield codes of linear codes. Some linear codes presented in this paper are optimal or almost optimal. To our knowledge,
many presented codes have new parameters. Specifically, the main work is summarized as follows:

\begin{description}
\item{$\bullet$} In Section $3$ and Section $4$,  we obtained the weight distribution of the subfield code $\mathcal{C}_k$, which is defined in~(\ref{code0}).

\item{$\bullet$} In section $5$,  the weight distribution of $\mathcal{\bar{C}}_k$ is determined,  where $\mathcal{\bar{C}}_k$ defined in~(\ref{eqbarC}) is a punctured code of  $\mathcal{C}_k$.

\item{$\bullet$} The parameters of the duals of $\mathcal{C}_k$ and $\mathcal{\bar{C}}_k$ are determined. The duals of $\mathcal{C}_k$ and $\mathcal{\bar{C}}_k$ are  optimal codes with respect to the Sphere Packing bound if $p>3$, and the dual of $\mathcal{\bar{C}}_k$ is optimal for the case $v_2(m)\leq v_2(k)$ if $p=3$ and $m>1$.

\item{$\bullet$} In Theorem \ref{theorem1}, the dual of $\mathcal{C}_k$ is a $p$-ary MDS code with parameters $[p+1,p-2,4]$ if $m=1$.

\item{$\bullet$} In Theorem \ref{thm3}, the dual of $\mathcal{\bar{C}}_k$ is a $p$-ary MDS code with parameters $[p,p-3,4]$ if $m=1$ and $p>3$.

\item{$\bullet$} Table \ref{Table5} demonstrated a class of $p$-ary MDS codes with parameters $[p, 3, p-2]$ if $m=1$.

\item{$\bullet$} Example \ref{example1}, Example \ref{example2} and Example \ref{example3} showed  some optimal or almost optimal codes with respect to the tables of best codes known maintained at http://www.codetables.de.
\end{description}

\end{document}